\documentclass[reqno,12pt]{article}
\usepackage[a4paper]{geometry}
\usepackage{amsmath,amsfonts,amssymb,amsthm,amstext,amscd,eucal,subfig,graphicx}
\usepackage[all]{xy}

\makeatletter \@addtoreset{equation}{section}

\makeatletter\renewcommand\section{\@startsection {section}{1}{\z@}%
                                   {-3.5ex \@plus -1ex \@minus -.2ex}%nn
                                   {2.3ex \@plus.2ex}%
                                   {\normalfont\large\bfseries}}
\renewcommand\subsection{\@startsection{subsection}{2}{\z@}%
                                     {-3.25ex\@plus -1ex \@minus -.2ex}%
                                     {1.5ex \@plus .2ex}%
                                     {\normalfont\bfseries}}

\newcommand{\be}{\begin{equation}}
\newcommand{\ee}{\end{equation}}
\newcommand{\beq}{\begin{eqnarray}}
\newcommand{\eeq}{\end{eqnarray}}
\newcommand{\bea}{\begin{eqnarray}}
\newcommand{\eea}{\end{eqnarray}}

\newcommand{\SO}{\mathrm{SO}}
\newcommand{\U}{\mathrm{U}}

\newcommand{\CC}{\mathbb{C}}

\newcommand{\ZZ}{\mathbb{Z}}

\def\CC{{\cal C}}

\def\CO{{\cal O}}

\def\sun2{$su(N)\times su(N)$}
\def\un2{$u(N)\times u(N)$}

\newcommand{\bbibitem}[1]{\bibitem{#1}\marginpar{#1}}

\def\Label#1{\label{#1}%
  \smash{\hbox to0pt{\raise1ex\hbox{\tiny[#1]}\hss}}}
\def\noLabels{\let\Label=\label}
\def\nobbibitem{\let\bbibitem=\bibitem}

\usepackage{hyperref}

\begin{document}

\noLabels % uncomment for final production
\nobbibitem % uncomment for final production

%%%%%%%%%%%%%%% TITLE PAGE %%%%%%%%%%%%%%%%%%%%

\begin{titlepage}

\begin{flushright}\vspace{-2cm}
{\small
{\tt Imperial/TP/2010/AD/01} \\
}\end{flushright} \vspace{12 mm}

%\vfil\
%%vfil

\begin{center}

{\Large{\bf The electrostatic view on M-theory LLM geometries}} \vfil \vspace{3mm}

{\large{{\bf Aristomenis Donos\footnote{e-mail:a.donos@imperial.ac.uk}$^{,a}$ and
Joan Sim\'on\footnote{e-mail:j.simon@ed.ac.uk}$^{,b}$}}}
\\

\vspace{8mm}

\bigskip\medskip
\begin{center} 
{$^a$ \it Theoretical Physics Group, Blackett Laboratory, \\
        Imperial College, London SW7 2AZ, U.K.}\\
{$^b$ \it School of Mathematics and Maxwell Institute for Mathematical Sciences,\\
King's Buildings, Edinburgh EH9 3JZ, United Kingdom}\\
\end{center}
\vfil

\end{center}
\setcounter{footnote}{0}

%%%%%%%%%%%%%%%%% ABSTRACT  %%%%%%%%%%%%%%%%%%%%%%%%%%%

\begin{abstract}
\noindent We describe the geometry of the $\mathbb{R}\times\SO(3)\times \SO(6)\times\U(1)$ invariant half-BPS M-theory configurations considered in LLM in terms of their electrostatic variables. We discuss both regular configurations, such as AdS${}_4\times$S${}^7$ and AdS${}_7\times$S${}^4$ vacua or simple excited solutions, and singular ones such as the superstar geometries. This allows us to identify  the appropriate boundary conditions describing the most general smooth and superstar-like singular configurations. We also compute their masses, matching the expected result from their microscopic interpretation, but now at finite radius of curvature.
\end{abstract}

%%%%%%%%%%%%%%%%%%%%%%%%%%%%%%%%%%%%%%%%%%%%%%%%%%%

\vspace{0.5in}

\end{titlepage}

%%%%%%%%%%%%% END OF TITLE PAGE %%%%%%%%%%%%%%%%%%%%%%%%%%

\renewcommand{\baselinestretch}{1.05}  %Line spacing

\tableofcontents

%\newpage

%%%%%%%%%%%%%%%%%% MAIN BODY %%%%%%%%%%%%%%%%%%%%%%%%%%

\section{Introduction}

The study of eleven dimensional $\mathbb{R}\times \SO(3)\times \SO(6)$ invariant half-BPS supergravity configurations  started in \cite{LLM}. They depend on a single scalar function satisfying a non-linear Toda equation. When there is an extra rotational $\U(1)$ isometry, there exists an implicit change of variables mapping the latter to a Laplace equation, and consequently, to an electrostatic problem. The purpose of this paper is to discuss the geometry of smooth and singular configurations with these symmetries at finite radius of curvature.
 
Unlike the case of $1/2$ BPS configurations in Type IIB with $\SO(4)\times \SO(4)$ symmetry where the precise boundary conditions imposed by regularity were explicitly spelled out in \cite{LLM}, a similar treatment of their eleven dimensional counterparts was still missing. There was no understanding of neither the electrostatic configuration responsible for the asymptotics nor the boundary conditions describing the physical solutions to the Laplace equation. In this paper we provide a precise answer to both questions, allowing us to state the precise boundary problem describing the most general smooth and singular configurations with these symmetries and to compute their masses. Just as smooth translationally invariant configurations have a finite discrete set of conducting disks as their building blocks \cite{LM}, the same will be true for the rotationally invariant ones.

First, we study the vacuum configurations. AdS${}_4\times$S${}^7$ is described by a semi-infinite line charge density and a single conducting disk attached to it. The electrostatic problem is that of solving Laplace's equation in the presence of a background potential, due to the line charge density, inducing some charge on the conducting disk, with the boundary condition that the potential at the disk is some given constant value. The latter is an integral equation with a unique solution found by Copson \cite{Copson}. The potential at the disk is fixed by regularity, whereas its size determines the AdS radius of curvature. AdS${}_7\times$S${}^4$ has a similar structure but instead of a finite conducting disk, it involves an infinite conducting plane. The electrostatic problem can be solved by the method of images. The distance between the line charge density and this plane determines the AdS${}_7$ radius of curvature. 

A given configuration allows an infinite number of equivalent electrostatic descriptions due to the existence of conformal invariance in the original LLM coordinates \cite{LLM}. We describe the effect that these conformal transformations have on the electrostatic data describing a single geometry.

Excited configurations are obtained either by moving the original conducting disk in the AdS${}_4\times$S${}^7$ or by adding extra finite conducting disks. We identify the set of coupled integral equations dealing with the appropriate boundary conditions to this problem, but we only discuss the simplest example for such excited configurations.

Even though we can not solve for the most general smooth configurations, we compute their masses. We do this by examining the electrostatic potential in a multipole expansion at large distances. Matching the asymptotic metric expansion with the one for the singular superstar configurations \cite{sugraref}, whose mass is well-known, we obtain
\begin{equation}
 M_4 =\frac{1}{2L_4}\,\sum_{i}\sum_{j=1}^{i}N_{2}^{j}N_{5}^{i}\,, \quad \quad M_7 =\frac{2}{L_7}\,\sum_{i}\sum_{j=1}^{i}N_{2}^{j}N_{5}^{i}\,.
\end{equation}
where $M_d$ stands for the mass in an asymptotic AdS${}_d$ with radius of curvature $L_d$. These expressions are the finite curvature gravity analogues of similar results obtained in field theory, either by analysing the spectrum of half-BPS configurations \cite{david,shahin-joan} in ABJM \cite{ABJM} or by considering the $N\to \infty$ limit in some matrix model dual formulations \cite{BMN} associated to taking the Penrose limit in the gravity description.

The singular type IIB superstar geometry was described in terms of a continuous distribution of rings and the origin of the singularity was interpreted in terms of coarse-graining of the quantum mechanical information characterising the quantum state in the fermionic phase space formulation of the system \cite{library,others}. We prove that the AdS${}_4\times$S${}^7$ and AdS${}_7\times$S${}^4$ superstars \cite{sugraref} are described in terms of continuous distributions of conducting disks. Any such distribution will describe a singular geometry, because the radial electric field component must vanish inside the continuous conductor. Geometrically, this forces both the 5-sphere and 2-sphere shrink to zero size at the same time in a non-smooth way, while the time coordinate becomes null. We develop a general boundary condition describing general superstar-like singular configurations in terms of the shape of the continuous boundary distribution of disks. The connection between singular geometries and continuous conductors was already pointed out in
\cite{mark-coarsegraining}. They used the statistical methods developed in \cite{library}, to identify singular geometries dual to thermal/typical states in different ensembles of half-BPS states. It would be interesting to identify a generic singular configuration in terms of typical states in ABJM. To achieve that one needs to understand the precise dictionary between the shape of the conductor and the properties of the quantum states in the field theory.

\section{Regular configurations}

Eleven dimensional half-BPS supergravity configurations invariant under an $\mathbb{R}\times\SO(3)\times\SO(6)$ isometry group are of the form \cite{LLM}
\begin{align}
ds_{11}^{2}= & e^{2\lambda}\left[ds_{4}^{2}+4\, d\Omega_{5}^{2}+y^{2}e^{-6\lambda}\, d\Omega_{2}^{2}\right]\label{eq:Metric_ansatz}\\
ds_{4}^{2}= & -4\,\left(y^{2}e^{-6\lambda}+1\right)\,\left(dt+v\right)^{2}+\frac{e^{-6\lambda}}{y^{2}e^{-6\lambda}+1}\,\left[dy^{2}+e^{D}\left(dx_{1}^{2}+dx_{2}^{2}\right)\right]\\
G_{4}= & F\wedge d\Omega_{2}\\
F= & d\left[-4y^{3}e^{-6\lambda}\left(dt+v\right)\right]+2\ast_{3}\left[e^{-D}y^{2}\left(\partial_{y}\frac{1}{y}\partial_{y}e^{D}\right)dy+y\partial_{i}\partial_{y}D\, dx^{i}\right]\\
e^{-6\lambda}=- & \frac{\partial_{y}D}{y\left(y\partial_{y}D-1\right)},\quad v_{i}=\frac{1}{2}\varepsilon_{ij}\partial_{j}D\end{align}
with the scalar function $D=D(x_i,\,y)$ $i=1,2$ satisfying the non-linear Toda equation
\begin{equation}
  \Box_{2}D+\partial_{y}^{2}e^{D}=0\,.
\label{eq:Toda}
\end{equation}

We will study the subset of $\U\left(1\right)$ invariant configurations under rotational symmetry in the $x_1-x_2$ plane. In polar coordinates $x_{1}+\imath\, x_{2}=x\, e^{\imath\beta}$, this corresponds to translations along $\beta$. Toda's equation \eqref{eq:Toda} gets mapped to the Laplace's equation \cite{Ward:1990qt}
\begin{equation}
  \frac{1}{\rho^{2}}\ddot{V}+V^{\prime\prime}=0\,,
\label{eq:Laplace}
\end{equation}
under the implicit change of variables
\begin{align}
  \rho^{2}= & x^{2}e^{D}\label{eq:trans1}\\
  y= & \rho\partial_{\rho}V=\dot{V}\label{eq:trans2}\\
  \log x= & \partial_{\eta}V=V^{\prime}\label{eq:trans3}
\end{align}
(From now on any dot will stand for $\rho\partial_p$ whereas any prime for $\partial_\eta$). After the coordinate transformation, the metric  \eqref{eq:Metric_ansatz} is
\begin{align}
ds_{11}^{2}= & \left(-\frac{\dot{V}\Delta}{2V^{\prime\prime}}\right)^{\frac{1}{3}}\left[4\, d\Omega_{5}^{2}-\frac{2V^{\prime\prime}\dot{V}}{\Delta}\, d\Omega_{2}^{2}+ds_{4}^{2}\right]\,,
\label{eq:metric} \\
ds_{4}^{2}= & -\frac{2V^{\prime\prime}}{\dot{V}}\left(d\rho^{2}+d\eta^{2}+\frac{\rho^{2}}{(\dot{V}^{\prime})^{2}-\ddot{V}V^{\prime\prime}}\, d\beta^{2}\right)-4\frac{(\dot{V}^{\prime})^{2}-\ddot{V}V^{\prime\prime}}{\Delta}\,\left(dt+v\right)^{2}\,,
\end{align}
where
\begin{equation*}
\Delta=(2\dot{V}-\ddot{V})V^{\prime\prime}+ (\dot{V}^{\prime})^{2}\,, \quad \quad
v = \left(1+\frac{\dot{V^\prime}}{\ddot{V}V^{\prime\prime}-(\dot{V^\prime})^2}\right)\,d\beta\,.
\end{equation*}

It was already argued in \cite{LLM} that the most general regular solution to \eqref{eq:Laplace} corresponds to a discrete distribution of conducting disks located at positions $\eta=\eta_i$ with radia $R_i$ and charge density $\sigma_i$. Each disk can carry a certain number of M2-branes coming from the flux quantisation condition \cite{LLM,LM}
\begin{equation}
N_{2} =\frac{1}{16\pi G_{11}T_{M_{2}}}\int_{C_{7}}\ast_{11}G_{4}\,, \\
\end{equation}
if there exists a smooth 7-cycle allowing us to compute the contribution from the i-th disk
\begin{align}
N_2^i & =\frac{1}{\pi^{3}\ell_{p}^{6}}\left.\int_{D_{i}}e^{D}dx_{1}dx_{2}\right|_{y=0}=\frac{2\cdot 2k}{\pi^{2}\ell_{p}^{6}}\int_{0}^{R_{i}}\rho^{2}\partial_{\rho}\left(\partial_{\eta}V_{\eta\rightarrow\eta_{i}^{+}}-\partial_{\eta}V_{\eta\rightarrow\eta_{i}^{-}}\right)d\rho\nonumber \\
& =-\frac{2\cdot 2k}{\pi^{3}\ell_{p}^{6}}Q_{i}\,, 
 \label{eq:M2_number}
\end{align}
with $Q_{i} =2\pi\int_{0}^{R_{i}}\sigma_{i}\,\rho d\rho$ being its charge. Each configuration can also carry M5-brane flux coming from the flux quantisation condition
\begin{equation}
N_{5} =\frac{1}{16\pi G_{11}T_{M_{5}}}\int_{C_{4}}G_{4}\,,
\label{eq:M5_number_d}
\end{equation}
if there exists a smooth 4-cycle in the geometry. One can construct one such 4-cycle for each segment between two consecutive disks located at $\eta_{i-1}=-d_{i-1}$ and $\eta_{i}=-d_{i}$. The number of M5-branes $N_5^i$ carried by such cycle is
\begin{equation}
N_{5}^{i} =\frac{1}{2\pi^{2}l_{p}^{3}}\left.\int_{D_{i}}y^{-1}e^{D}dx_{1}dx_{2}\right|_{y=0}=\frac{2k}{\pi \ell_{p}^{3}}\left.\int_{d_{i+1}}^{d_{i}}\frac{\rho}{\partial_{\rho}V}\partial_{\eta}^{2}V\right|_{\rho=0}=\frac{2\cdot 2k}{\pi \ell_{p}^{3}}\left(d_{i}-d_{i-1}\right)\,.
\label{eq:M5_number}
\end{equation}
To compute both flux integrals we parameterised the range of the compact variable $\beta$ as $\beta\sim \beta + 4k\pi$.  We will review the construction of the 4-cycles and 7-cycles appearing in these integrals in subsection 2.5 and postpone a discussion on the actual values that $k$ can take until we have solved the electrostatics problem associated with the two vacuum configurations $AdS_4\times S^7$ and $AdS_7\times S^4$. 

One important feature of the metric \eqref{eq:Metric_ansatz} is its invariance under general conformal transformations \cite{LLM}
\begin{equation}
  \omega_1\equiv x_1+\imath\,x_2 \to f(\omega_1)\,, \quad \quad D\to D - \log \left|\frac{df}{d\omega_1}\right|^2\,.
\label{eq:conft}
\end{equation}
We will later comment on the action that the subset of these transformations preserving the $\U(1)$ invariance along $\partial_\beta$ have in the electrostatic description of these configurations.

\subsection{The $AdS_{4}\times S^{7}$ ground state\label{AdS4_ground}}

The metric of global $AdS_{4}\times S^{7}$ is 
\begin{equation}
L_4^{-2}\,ds_{11}^{2}=-\left(1+r^{2}\right)\, d\tau^{2}+r^{2}\, d\Omega_{2}^{2}+\frac{1}{1+r^{2}}dr^{2}+4\,\left(\cos^{2}\theta\, d\phi^{2}+d\theta^{2}+\sin^{2}\theta\, d\Omega_{5}^{2}\right)\,.
\label{eq:ads4}
\end{equation}
where the $AdS_4$ radius of curvature $L_4=\ell_p\,\left(\pi^2\,N_2/2\right)^{1/6}$ is given in terms of an integer $N_2$.

Due to the existence of the conformal transformations \eqref{eq:conft}, we know there must exist an infinite number of ways of describing \eqref{eq:ads4} in the original LLM coordinates $\{x,\,\beta,\,t\}$. On the other hand, the product of the 2-sphere and 5-sphere radia fixes the $y$ coordinate uniquely
\begin{equation}
  y =L_4^{3}\,r\,\sin^{2}\theta\,.
\end{equation}
The technical origin of this multiplicity can be seen, for example, from the matching of the $r-\theta$ and $y-x$ planes in \eqref{eq:ads4} and in LLM coordinates. The latter does not determine the scalar function $e^D$, which remains entirely undetermined. Once a particular $e^D$ is chosen, it determines $x$, and from that, the choice of $\U(1)$'s is fixed, modulo discrete transformations in $\{\tau,\,\phi\}$. For example, one can check that the choices  
\begin{align}
x =L_4^{3}\,\sqrt{1+r^{2}}\,\cos^{2}\theta\,, \quad & \quad
e^{D} =\tan^{2}\theta\,, \label{eq:xDmatch} \\
d\beta = 2d\phi+d\tau\,, \quad & \quad
dt = -d\phi \label{eq:time_scaling}
\end{align}
are a consistent possibility. The same is true for
\begin{equation}
x = \left(1 + \frac{r^2}{4}\right)^{1/4}\,\cos\theta\,, \quad \quad
e^D = 4L_4^{6}\sqrt{1+\frac{r^2}{4}}\,\sin^2\theta\,.
\end{equation}
The two mappings are related through the conformal transformation $z_1=L_4^3\,\left(\omega_1\right)^2$ where 
\begin{equation}
  |z_1|=\sqrt{1+r^2}\,\cos^2\theta \quad \quad \text{and} \quad \quad |\omega_1|=(1+r^2/4)^{1/4}\,\cos\theta\,.
\end{equation}
This conformal transformation fixes the relation between both angular variables, and consequently also explains the relation among both $\U(1)$'s. In particular, if $z_1=\hat{x}\,e^{i\hat\beta}$ and $\omega_1=x\,e^{i\beta}$, then it is clear $\hat\beta\sim \hat\beta + 4\pi$, whereas $\beta\sim \beta + 2\pi$, since $\hat\beta=2\beta$. Thus, different conformal frames correspond to different $\U(1)$ embeddings and consequently to different values of the constant $k$, even though the full flux integrals are conformally invariant. Clearly, by considering conformal transformations of the form $z_1\propto(\omega_1)^n$, which preserve the $\U(1)$ invariant nature of the initial angular variable $\beta$, we could generate an infinite number of different mappings indexed by the integer $n$. Different mappings would correspond to different values of the constant $k$ parameterising the range of the LLM compact coordinate $\beta$, though all of them would describe the same number of M2-branes $N_2$.

Next, we would like to understand the electrostatic configuration giving rise to $AdS_4\times S^7$. Using \eqref{eq:trans1} and \eqref{eq:xDmatch}  fixes $\rho$ to be
\begin{equation}
  \rho =\frac{L_4^{3}}{2}\sqrt{1+r^{2}}\sin2\theta\,.
\end{equation}
Requiring the absence of cross-terms in the $\rho-\eta$ plane fixes
\begin{equation}
\eta=-\frac{L_4^{3}}{2}\, r\cos2\theta\,.
\end{equation}

Inverting these relations
\begin{align}
  r^{2}= \frac{1}{2L_4^{6}}\left(-L_4^{6}+4(\eta^{2}+\rho^{2})+C\right)\,, \quad & \quad
  \cos2\theta= -\frac{\eta}{\sqrt{2}L_4^{3}\left|\eta\right|}\,\left(L_4^{6}-4(\eta^{2}+\rho^{2})+C\right)^{\frac  {1}{2}}\,, \nonumber \\
  C= & \sqrt{16L_4^{6}\eta^{2}+\left(4\eta^{2}+4\rho^{2}-L_4^{6}\right)^{2}}\,,
\end{align}
allows us to relate both types of LLM coordinates as follows
\begin{align}
  y=\rho\partial_{\rho}V= & \frac{1}{2\sqrt{2}}\left(-L_4^{6}+4\eta^{2}+4\rho^{2}+C\right)^{\frac{1}{2}}+\eta\label{eq:pr_V}\\
\log x=\partial_{\eta}V= & \log\left[\frac{1}{2\sqrt{2}}\left(L_4^{6}+4\eta^{2}+4\rho^{2}+C\right)^{\frac{1}{2}}\,\left(1-\frac{\eta}{\sqrt{2}L_4^{3}\left|\eta\right|}\,\left(L_4^{6}-4\eta^{2}-4\rho^{2}+C\right)^{\frac{1}{2}}\right)\right]\label{eq:pn_V}
\end{align}

\paragraph{The $AdS_4\times S^7$ electrostatics :} One can check that the consistency relation
$ \partial_{\eta}y=\rho\partial_{\rho}\log x$ holds everywhere, i.e. $\rho\partial_{\rho\eta} V = \rho\partial_{\eta\rho} V$. Furthermore, Laplace's equation \eqref{eq:Laplace}
\begin{equation}
  \frac{1}{\rho}\partial_{\rho}y+\partial_{\eta}\log x=0
\end{equation}
also holds, but only away from certain regions where charge sources exist. This is what we study next.

First, notice the electric field $\partial_{\rho}V$ in \eqref{eq:pr_V} diverges at $\eta>0$ and close to the axis $\rho=0$. This indicates the presence of a semi-infinite line charge distribution $\lambda\left(\eta\right)$ given by
\begin{equation}
  \lambda\left(\eta\right) = \rho\partial_{\rho}V \rightarrow\begin{cases}
  0 & ,\eta<0\\
  2\eta & ,\eta>0\end{cases}
 \label{eq:lcharge1}
\end{equation}

Second, the same electric field $\partial_\rho V$ vanishes along the disk located at $\eta=0$ and with radius $R=\frac{L_4^{3}}{2}$, i.e. described by $\rho<\frac{L_4^{3}}{2}$. This defines a conducting disk whose surface charge distribution $\sigma (\rho)$ can be computed using the discontinuity of the derivative \eqref{eq:pn_V} across $\eta=0$ for $\rho<L_4^{3}/2$ due to the factor $\frac{\eta}{\left|\eta\right|}$
\begin{align}
\partial_{\eta}V_{\eta\rightarrow0^{+}}=\log\left[\frac{L_4^{3}}{2}\left(1-\sqrt{1-4L_4^{-6}\rho^{2}}\right)\right]\,,\quad & \quad
\partial_{\eta}V_{\eta\rightarrow0^{-}} =\log\left[\frac{L_4^{3}}{2}\left(1+\sqrt{1-4L_4^{-6}\rho^{2}}\right)\right]\,, \nonumber\\
\sigma(\rho) = \partial_{\eta}V_{\eta\rightarrow0^{+}}-\partial_{\eta}V_{\eta\rightarrow0^{-}} & =\log\left[\frac{1-\sqrt{1-4L_4^{-6}\rho^{2}}}{1+\sqrt{1-4L_4^{-6}\rho^{2}}}\right]\,. 
\label{eq:vac_surface_charge}
\end{align}

To sum up, the geometry of the $AdS_4\times S^7$ vacuum is encoded in terms of a semi-infinite line charge distribution $\lambda(\eta)$ \eqref{eq:lcharge1} at $\rho=0$ and a conducting disk located at $\eta=0$ of size $R=\frac{L_4^{3}}{2}$ with surface charge distribution $\sigma(\rho)$ \eqref{eq:vac_surface_charge}. 

We can map this information to the LLM description through the
coordinate transformations \eqref{eq:trans2} and \eqref{eq:trans3}. The line charge distribution at
$\rho=0$, $\eta>0$ is mapped to
\begin{equation}
x =0\,, \qquad \qquad
y =2\eta\,.
\end{equation}
Thus, as we move along the line distribution towards smaller values of $\eta$, we move along the axis $x=0$ towards the plane located at $y=0$. The regions $\eta\to 0^\pm$ close to the conducting disk are mapped to
\begin{equation}
y =0\,, \quad \quad
x =\frac{L_4^{3}}{2}\left(1\mp \sqrt{1- 4L_4^{-6}\rho^{2}}\right)\,.
\end{equation}
Thus, as we move along the disk for $\eta\to 0^+$, $x$ increases from 0 to $L_4^3/2$. If we move along the disk for $\eta\to 0^-$, $x$ increases from $L_4^3/2$ to $L_4^3$ as we move from the rim towards the disk axis at $\rho=0$. Finally, the segment at $\eta<0$, $\rho=0$ where $\lambda(\eta)$ vanishes is mapped to
\begin{equation}
y =0\,, \qquad \qquad
x =L_4^{3}\sqrt{1+4L_4^{-6}\eta^{2}}\,.
\end{equation}
It is an interesting observation that the $x=0$ axis corresponds to the set of points where the $\U(1)$ circle shrinks smoothly.
As soon as we reach the $y=0$ plane, the region $x\in [0,\, L_4^3]$ corresponds to the set of points where the 2-sphere shrinks smoothly, whereas the region $x>L_4^3$ describes the points where the 5-sphere collapses smoothly. This structure is reminiscent of the one reported for 1/2 BPS configurations in type IIB \cite{LLM}.

\paragraph{First principles derivation:} Given the nonlinearity of the Toda equation, it is convenient to rederive the previous results by solving the Laplace equation in the presence of a line charge distribution and a conducting disk before attempting to solve more complicated electrostatic configurations.

In general, a potential $V$ reproducing \eqref{eq:pr_V} and \eqref{eq:pn_V} will satisfy Laplace's equation in the presence of some charge distribution $q(\eta,\,\rho)$
\begin{equation}
  \frac{1}{\rho}\partial_{\rho}\left(\rho\partial_{\rho}V\right)+\partial_{\eta}^{2}V=\sigma\left(\rho,\eta\right)\,.\label{eq:Laplace_sourced}
\end{equation}
Thus, its general solution is
\begin{equation}
  V=-\frac{1}{4\pi}\int d^{3}\mathbf{x}^{\prime}\,\frac{\sigma\left(\mathbf{x}^{\prime}\right)}{\left|\mathbf{x}-\mathbf{x}^{\prime}\right|}
\end{equation}

We can rephrase the current electrostatic problem as follows : we have the line charge distribution $\lambda(\eta)$ \eqref{eq:lcharge1} and a conducting disk at $\eta=0$ of radius $R=\frac{L_4^3}{2}$. $\lambda(\eta)$ produces a potential $V_b$, which we will think of as a background potential in which the conducting disk is placed. As a result, there will be some induced surface charge distribution $\sigma(\rho)$ on the disk, producing a potential $V_d$ so that the total electrostatic potential equals
\begin{equation}
  V=V_{d}+V_{b}\,.
\end{equation}

The background potential generated by the semi-infinite line charge distribution $\lambda(\eta)$ is formally given by 
\begin{equation}
  V_{b}=-\frac{1}{2}\int_{0}^{\infty}\frac{2\eta^{\prime}}{\sqrt{\rho^{2}+\left(\eta-\eta^{\prime}\right)^{2}}}\, d\eta^{\prime}\,.
\end{equation}
This integral diverges and requires regularization. We will use the freedom to shift the potential\footnote{The origin of this symmetry is the existence of the conformal symmetry in the initial LLM coordinates. See subsection 2.3 for further comments.} by linear terms in $\eta$ to achieve a finite background potential
\begin{align}
  V_{b} & =-\lim_{L\rightarrow+\infty}\left[\frac{1}{2}\,\int_{0}^{L}\frac{2\eta^{\prime}}{\sqrt{\rho^{2}+  \left(\eta-\eta^{\prime}\right)^{2}}}\, d\eta^{\prime}+L+\eta\,\ln\left(2L\right)\right]\nonumber \\
 & =\sqrt{\eta^{2}+\rho^{2}}+\eta\,\log\left[-\eta+\sqrt{\eta^{2}+\rho^{2}}\right]\,.
 \label{eq:bpotential4}
 \end{align}
Notice this expression still satisfies \eqref{eq:Laplace_sourced}. 

The disk potential $V_d$ is determined by the induced surface charge distribution $\sigma(\rho)$
\begin{equation}
  V_{d}=-\frac{1}{4\pi}\int_{0}^{R}\rho^{\prime}d\rho^{\prime}\int_{0}^{2\pi}d\phi^{\prime}\,\frac{\sigma\left(\rho^{\prime}\right)}{\sqrt{\rho^{2}+\rho^{\prime2}-2\rho\rho^{\prime}\cos\phi^{\prime}+\eta^{2}}}\,,
\end{equation}
where we already used the fact that we are dealing with $\U(1)$ invariant configurations.

Our electrostatic problem defines a boundary problem since the total potential must be a constant $V_{0}$ on the surface of the conducting disk
\begin{equation}
  V_{0} =V_{b}\left(\rho,0\right)+V_{d}\left(\rho,0\right),\quad\rho<R=\frac{L_4^3}{2}\,.
\label{eq:boundaryproblem}
\end{equation}
Notice this condition defines an integral equation for the disk charge density $\sigma(\rho)$. 

This electrostatic problem was solved by Copson \cite{Copson}. Specifically, he solved the electrostatic problem of a conducting disk immersed in some external potential satisfying the boundary condition of keeping the potential at the disk equal to some continuous and differentiable function $f(\rho)$. In our situation,
\begin{equation}
  V_d(\rho,\,0) = V_0 - \rho = f(\rho)\,,
\end{equation}
Copson's {\it unique} solution is given by
\begin{align}
  S\left(\rho\right) & =\frac{1}{2\pi}\,\frac{\partial}{\partial\rho}\int_{0}^{\rho}\frac{tf\left(t\right)}{\sqrt{\rho^{2}-t^{2}}}dt\,,  \label{eq:Copson_S}\\
  \sigma\left(\rho\right) & =\frac{8}{\rho}\,\frac{\partial}{\partial\rho}\int_{\rho}^{R}\frac{tS\left(t\right)}{\sqrt{t^{2}-\rho^{2}}}dt\,.
\label{eq:Copson_sigma}
\end{align}
Computing the integrals, we obtain
\begin{align}
  S\left(\rho\right) & =\frac{V_{0}}{2\pi}-\frac{1}{4}\rho\\
  \sigma\left(\rho\right) & =-2\,\log\left(\frac{R+\sqrt{R^{2}-\rho^{2}}}{\rho}\right)-\left(\frac{4V_{0}}{\pi}-2R\right)\,\frac{1}{\sqrt{R^{2}-\rho^{2}}}
\end{align}
Regularity will {\it not} allow an arbitrary disk potential value $V_0$ but fixes it to be
\begin{equation}
  V_{0}=\frac{\pi R}{2}\,.
\end{equation}
It is reassuring that this fact guarantees the vanishing of the surface charge density $\sigma(\rho)$ at the rim of the disk $\rho=R$ and as a result, the continuity of the first derivative $\partial_{\eta}V$ there. As we will discuss later, the observation that regularity determines the value of the potential at the disk seems to be generic.

Our first principle derivation assumed the disk was located at $\eta=0$, but worked for any disk radius $R$. Obviously, this last parameter will determine its charge $Q$
\begin{equation}
  Q = 2\pi\int_0^R \sigma(\rho)\,\rho\,d\rho = -2\pi\,R^2\,.
\end{equation}
Substituting this charge in the flux quantisation condition \eqref{eq:M2_number} fixes
$R=\frac{L_4^{3}}{2}$. With this choice, we both reproduce the surface charge density \eqref{eq:vac_surface_charge} computed by the direct matching of metrics and match the quantisation condition \eqref{eq:M2_number} telling us the solution carries $N_2$ units of M2-brane flux.

\subsection{The $AdS_7\times S^4$ ground state}\label{AdS7_Ground}

The ansatz \eqref{eq:Metric_ansatz} can also describe the $AdS_{7}\times S^{4}$ vacuum 
\begin{equation}
ds_{11}^{2}=L_7^{2}\left[-\left(1+r^{2}\right)\, d\tau^{2}+r^{2}\, d\Omega_{5}^{2}+\frac{1}{1+r^{2}}dr^{2}+\frac{1}{4}\left(\cos^{2}\theta\, d\phi^{2}+d\theta^{2}+\sin^{2}\theta\, d\Omega_{2}^{2}\right)\right]\,
\end{equation}
where $L_7 = 2\ell_p\,(\pi N_5)^{1/3}$ is the $AdS_7$ radius of curvature.

The matching of this geometry to the original LLM coordinates shares most of the comments mentioned in the previous section. As an example of the infinite set of such mappings, we can give
\begin{align}
y= \frac{L_7^3}{8}\,r^{2}\,\sin\theta , \quad \quad x&=\frac{L_7^{3}}{8}\left(1+r^{2}\right)\cos\theta , \quad \quad
e^{D}= \frac{r^{2}}{1+r^{2}}\,,\\
d\tau=dt ,  & \quad \quad d\beta = -2\,dt + d\phi\,.
\label{eq:matchad7}
\end{align}
In this particular mapping, $\beta\sim \beta + 2\pi$. Consequently, the parameter $k$ entering in our flux quantisation conditions will equal $k=1/2$.

Let us describe this vacuum geometry in the electrostatics coordinates. Using \eqref{eq:trans1} and analysing the mapping from the $r-\theta$ plane to the $\rho-\eta$ plane we can fix the transformation to be
\begin{equation}
\rho= \frac{L_7^{3}}{8}\,r\left(1+r^{2}\right)^{\frac{1}{2}}\cos\theta\,, \quad \quad
\eta= \frac{L_7^{3}}{8}\left(\frac{1}{2}+ r^{2}\right)\sin\theta\,.
\end{equation}
Inverting these,
\begin{align}
\sin\theta & =\frac{8}{\sqrt{2}L_7^{3}}\left(\frac{L_7^{6}}{2^6}+4\left(\eta^{2}+\rho^{2}\right)-C^{\frac{1}{2}}\right)^{\frac{1}{2}}\,,\\
\cos\theta & =\frac{8}{\sqrt{2}L_7^{3}}\left(\frac{L_7^{6}}{2^6}-4\left(\eta^{2}+\rho^{2}\right)+C^{\frac{1}{2}}\right)^{\frac{1}{2}}\,,\\
r^{2} & =-\frac{1}{2}+\frac{4}{\sqrt{2}L_7^{3}}\left(\frac{L_7^{6}}{2^6}+4\left(\eta^{2}+\rho^{2}\right)+C^{\frac{1}{2}}\right)^{\frac{1}{2}}\,,\\
C & =\frac{L_7^{6}}{4}\rho^{2}+\left(\frac{L_7^{6}}{2^6}-4\left(\eta^{2}+\rho^{2}\right)\right)^2\,.
\end{align}
We now use these expressions along with \eqref{eq:trans2} and \eqref{eq:trans3}
to learn about the electrostatic potential
\begin{align}
\rho\partial_{\rho}V= & y=\eta-\frac{1}{2\sqrt{2}}\left(\frac{L_7^6}{2^6}+4\left(\eta^{2}+\rho^{2}\right)-C^{\frac{1}{2}}\right)^{\frac{1}{2}}\,.\label{eq:pr_V2}\\
\partial_{\eta}V= & \ln x=\ln\left[\frac{1}{2\sqrt{2}}\left(1+\frac{2^3}{\sqrt{2}L_7^{3}}\left(\frac{L_7^{6}}{2^6}+4\left(\eta^{2}+\rho^{2}\right)+C^{\frac{1}{2}}\right)^{\frac{1}{2}}\right)\left(\frac{L_7^{6}}{2^6}-4\left(\eta^{2}+\rho^{2}\right)+C^{\frac{1}{2}}\right)^{\frac{1}{2}}\right]\,.
\label{eq:pn_V2}
\end{align}
As long as $\eta>0$ or $\{\rho>0\,, \eta>\frac{L_7^{3}}{16}\}$, the above potential
satisfies both Laplace's equation and the consistency condition
\begin{align}
\frac{1}{\rho}\partial_{\rho}y+\partial_{\eta}\log x &=0\,, \nonumber \\
\partial_{\eta}y &=\rho\partial_{\rho}\log x\,.
\end{align}

As for the $AdS_4\times S^7$ configuration, we now examine the location of sources.
First, consider the electric field $\partial_\rho V$. From \eqref{eq:pr_V2}, 
we see that at $\eta\to 0^{+}$ it vanishes
\begin{equation}
\partial_{\rho}V_{\eta\to 0^{+}}=0\,, \quad \quad \forall\,\rho\,.
\end{equation}
This points out the existence of an infinite conducting surface at $\eta=0$. Its charge surface density $\sigma(\rho)$ can be computed from the discontinuity in the electric field $\partial_\eta V$ at the conducting plane
\begin{equation}
  \sigma = \partial_\eta V_{\eta\to0^+}- \partial_\eta V_{\eta\to0^-} = \log\left[\frac{L_7^{3}}{16}+\sqrt{\frac{L_7^{6}}{2^8}+\rho^{2}}\right]\,.
\label{eq:sigma2}  
\end{equation}
Furthermore, at $\rho\to 0$ and $\eta>\frac{L_7^{3}}{16}$
\begin{equation}
\rho\partial_{\rho}V_{\rho\to 0}=\eta - \frac{L_7^3}{16}\,.
\end{equation}
Thus, the electric field diverges in this region, and as for the $AdS_4\times S^7$, we will interpret this as due to the presence
of a semi-infinite line charge distribution
\begin{equation}
\lambda\left(\eta\right)=\begin{cases}
0 & ,0<\eta<\frac{L_7^{3}}{16}\\
\eta-\frac{L_7^{3}}{16} & ,\eta>\frac{L_7^{3}}{16}
\end{cases}
\label{eq:lcharge2}
\end{equation}
located at $\rho=0$.

To sum up, the geometry of the $AdS_7\times S^4$ vacuum is encoded in terms of a semi-infinite line charge distribution $\lambda(\eta)$ \eqref{eq:lcharge2} at $\rho=0$ and an infinite conducting plane located at $\eta=0$  with surface charge density $\sigma(\rho)$ \eqref{eq:sigma2}.

\paragraph{First principle derivation:} Ignoring the infinite plane, the background potential due to \eqref{eq:lcharge2} and after appropriate regularization would equal :
\begin{align}
  V_b(\rho,\eta) &=\frac{1}{2}\left[\sqrt{\rho^2+\left(\eta-\frac{L_7^3}{16}\right)^2} \right. \nonumber \\
  & \left. + \left(\eta-\frac{L_7^3}{16}\right)\log\left[-\left(\eta-\frac{L_7^3}{16}\right) + \sqrt{\rho^2 + \left(\eta-\frac{L_7^3}{16}\right)^2}\right]\right]\,.
\label{eq:ads7backpot}
\end{align}
Due to the presence of the infinite conducting plane at $\eta=0$, we must modify the total potential to keep the value of the potential at the plane fixed. This can be accomplished by using the method of images. Thus, the total potential equals
\begin{equation}
  V =\begin{cases} V_b(\rho,\eta) - V_b(\rho,-\eta) + V_0, &\eta>0\\ V_{0},& \eta<0 \end{cases},
 \label{eq:tads7pot}
\end{equation}
where $V_0$ is the potential at such infinite plane. The potential $V$ is constant below the conducting plane $\eta<0$ since that region is screened. Notice the perpendicular component of the electric field $\partial_{\eta}V$ is discontinuous along the $\eta=0$ plane yielding an induced charge density
\begin{equation}
\sigma=\partial_{\eta}V_{\eta\to 0^{+}}-\partial_{\eta}V_{\eta\to 0^{-}}=\log\left[\frac{L_7^{3}}{16}+\sqrt{\frac{L_7^{6}}{2^8}+\rho^{2}}\right]\,,
\end{equation}
matching \eqref{eq:sigma2}.

\begin{figure}
\centering
\includegraphics{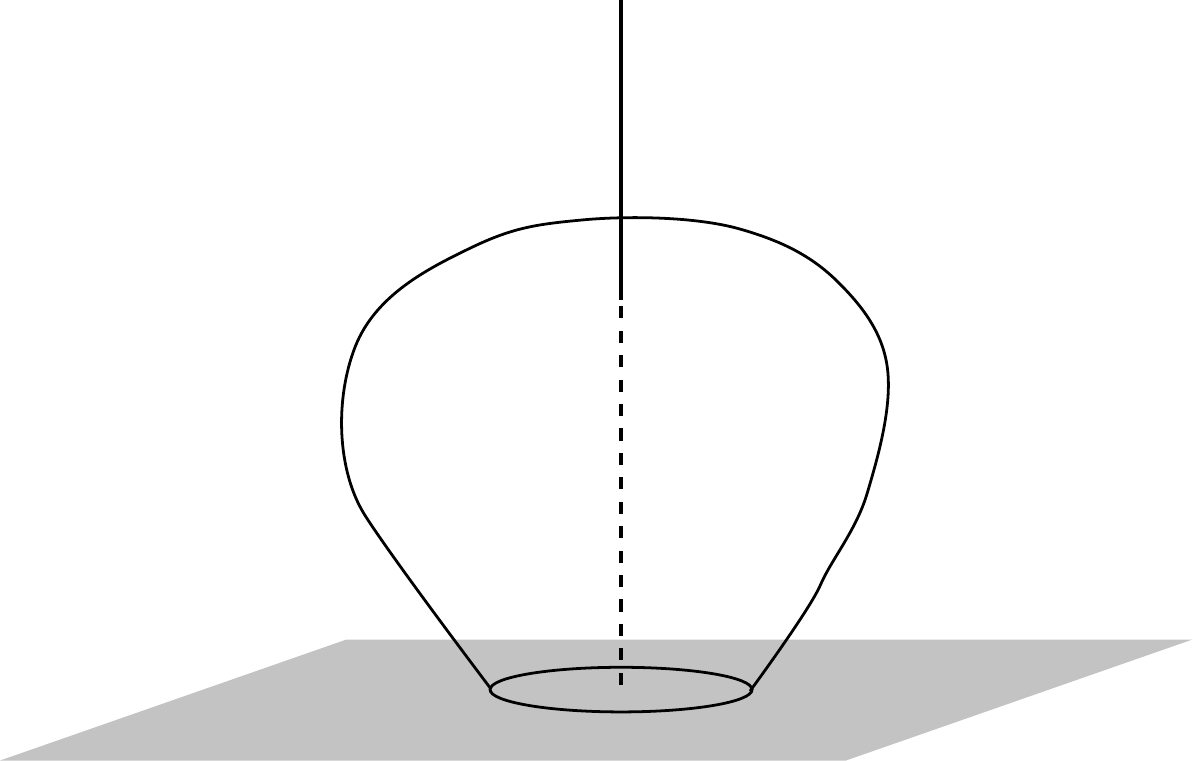}
\caption{Surface which when fibered over the round $S^{2}$ gives the unique four-cycle of the $AdS_{7}\times S^{4}$ background.}
\label{AdS7FC}
\end{figure}

One might think this non-zero charge distribution implies the existence of non-trivial M2-brane charge. But the only available cycle in our geometry is the four-cycle appearing in figure \ref{AdS7FC}.
Thus, $AdS_4\times S^7$ carries no M2-brane flux, whereas its M5-brane flux will only depend on the distance between the line and plane charge densities
\begin{equation}
  N_5 = \frac{2\cdot 2k}{\pi\,\ell_p^3}\,\frac{L_7^3}{2^4}\,.
\end{equation}
This reproduces the appropriate quantisation condition when $k=1/2$, which is the right value for the embedding used in \eqref{eq:matchad7}, since $L_7=2\ell_p\left(\pi N_5\right)^{1/3}$.

\subsection{Conformal symmetry action in the electrostatic variables}

Both vacuum configurations are described in terms of  semi-infinite line charge distributions and finite (infinite) conducting charged disks. Given the existence of the conformal symmetry \eqref{eq:conft}, it is natural to wonder what the action of the subset of such transformations preserving the $\U(1)$ invariance in our configurations is in their electrostatic description. Consider one such transformation
\begin{equation}
  z=\lambda\,\omega^n
\label{eq:conf}
\end{equation}
with $\lambda=|\lambda|\,e^{i\gamma}\in\CC$, $n\in \ZZ$, $z=\hat{x}e^{i\hat\beta}$ and $\omega=x\,e^{i\beta}$. By definition
\begin{equation}
  \hat{x} = |\lambda|\,x^n\,, \quad\quad \hat\beta=n\beta + \gamma\,.
\end{equation}
Using the implicit change of coordinates \eqref{eq:trans1}-\eqref{eq:trans3} in both conformally related frames, we derive the relation
\begin{equation}
  \partial_{\hat\eta}\hat{V} = n\,\partial_\eta V + \log|\lambda|\,,
\end{equation}
where $\hat{V}$ stand for the electrostatic potential in the z-frame. Conformal invariance of the metric \eqref{eq:conft}  implies
\begin{equation}
  e^{\hat{D}} = \frac{e^D}{n^2\,|\lambda|^2\,x^{2(n-1)}}\,,
\end{equation}
allowing us to conclude $\hat\rho = \rho/n$. Thus, disk sizes $R_i$ are not invariant, but transform under these transformations
\begin{equation}
\hat{R}_i = \frac{R_i}{n}\,.
\end{equation}
Requiring Laplace's equation \eqref{eq:Laplace} to transform homogeneously, we learn
\begin{equation}
  \frac{\partial\eta}{\partial\hat\eta}=n\,,
 \label{eq:framet}
\end{equation}
where we used $\partial^2_{\hat\eta}\hat{V} = n\,\frac{\partial\eta}{\partial\hat\eta}\,\partial^2_\eta V$. 

A first consequence of \eqref{eq:framet} is the non-invariance of the semi-infinite line charge distribution since the latter was linear in $\eta$. Secondly, the potentials in both frames are related as
\begin{equation}
  \hat{V} = V + \frac{\eta}{n}\log|\lambda| = V + \hat\eta\,\log|\lambda|\,.
\end{equation}
Notice that for $n=1$, this is precisely the symmetry transformation that we used to regularise the background potentials in both vacuum configurations. Thirdly, since $\rho\partial_\rho$ is invariant, $y$ is invariant, as expected.

Since the Laplace equation transforms, in the presence of charge density sources $\sigma_i$, these will transform. If the latter are localised at $\eta=\eta_i$ and since delta functions transform under the conformal transformation, we conclude
\begin{equation}
  \hat\sigma_i = n\,\sigma_i\,.
\end{equation}
This transformation law is consistent with the invariance of the flux integrals. Consider the M5-brane flux  \eqref{eq:M5_number}. If $\beta\sim\beta + 4\pi k$, then $\hat{k}=n\,k$. Thus, the product $k\eta_i$ is invariant. Furthermore, the relation between $\eta$ and $\hat\eta$ allows a constant shift, but the latter leaves the M5-brane flux invariant since this quantity only depends on the differences $\eta_i-\eta_{i-1}$. Consider the M2-brane flux  \eqref{eq:M2_number}. Its invariance requires $Q_i=n\hat{Q}_i$. This is consistent with our transformation laws since
\begin{equation}
  n\hat{Q}_i = n\int \hat\sigma\hat\rho\,d\hat\rho = \int \sigma\rho\,d\rho = Q_i\,.
\end{equation}

Thus, the electrostatic data characterising a given smooth $\U(1)$ invariant configuration is not invariant under $\U(1)$ preserving conformal transformations. In particular, the slope of the semi-infinite line charge distribution, the location, size and charge density of the different discrete disks are not invariant, but the individual quantised fluxes are, and so their masses will be.

\paragraph{AdS${}_{7}\times$S${}^{4}$ as a limit of AdS${}_{4}\times$S${}^{7}$ :} The electrostatic descriptions for both vacua are rather similar. They both involve a semi-infinite line charge density and a conducting surface : a disk of finite size for AdS${}_{4}\times$S${}^{7}$ and an infinite plane for AdS${}_{7}\times$S${}^{4}$. It is natural to wonder whether one can reproduce the second geometry from an appropriate infinite charge limit of the first.

Since the embeddings we used for both geometries have different $\U(1)$ periodicities, i.e. different values of $k$, we first apply a conformal transformation to match these. This will modify both the slope of the electrostatic and the charge densities according to
\begin{equation}
  \lambda(\eta)\to k\,\lambda(\eta)\,, \quad \quad \sigma(\rho)\to k\,\sigma(\rho)\,.
\end{equation}
Furthermore, the location of the semi-infinite line distribution in AdS${}_{7}\times$S${}^{4}$ is shifted from the one in the AdS${}_{7}\times$S${}^{4}$. Thus, we also need to consider the shift $\eta\to\eta-L_7^3/16$, which is compatible with \eqref{eq:framet}. After these transformations, taking the $R\to \infty$ in the AdS${}_{4}\times$S${}^{7}$ configuration, the disk charge density behaves like
\begin{equation}
\sigma=-\,\log\left(2R\right)+\log\left(\frac{L_7^{3}}{16}+\sqrt{\frac{L_7^{6}}{2^8}+\rho^{2}}\right)\,.
\label{eq:AdS7_distr_limit}
\end{equation}
Notice the finite piece reproduces the $AdS_7\times S^4$ charge density \eqref{eq:sigma2}.
As for the divergent piece, it modifies the total potential by
\begin{equation}
V=-\frac{1}{2}\ln\left(2R\right)\eta+V_{reg}\,,
\end{equation}
where $V_{reg}$ stands for the AdS${}_{7}\times$S${}^{4}$ potential. The linear divergent term can be absorbed by an appropriate $|\lambda|$ rescaling in \eqref{eq:conf}.

Thus, there exists a formal connection between both vacuum configurations and the existence of conformal transformations makes it more precise. Notice the above construction can be done for any conducting disk located at an arbitrary $\eta_0$. After the charge is sent to infinity, we would only keep the region $\eta>\eta_{0}$ as part of our spacetime.

\subsection{Regularity\label{sec:Regularity}}

In previous sections, we developed the geometry of the electrostatic problem describing two different vacuum configurations. The latter involve semi-infinite line charge densities and conducting disks. In this section we examine the regularity of the metric \eqref{eq:metric} in regions close to these disks where various derivatives of the potential $V$ take their zeros or become infinite close to their rims. We also examine the behaviour of the metric close to the $\rho=0$ axis where the linear charge distribution is non-zero. The generic electrostatic configuration we imagine having is a line charge
distribution at $\rho=0$ 
\begin{equation}
\lambda\left(\eta\right)=\begin{cases}
2k\eta & ,\eta>0\\
0 & ,\eta<0
\end{cases}
\label{eq:glinear}
\end{equation}
and $N$ circular conductors of radii $R_{i}$ at $\eta_{i}<0$ centered
at $\rho=0$. 

In the following paragraphs we focus on four different regions of the $\left(\eta,\rho\right)$ space.
First, we show the region $\rho=0$, $\eta>0$ where the line charge distribution is reached is smooth. Second,
we study the semi-infinite axis $\rho=0$, $\eta<0$, $\eta\neq\eta_{i}$ and show the 5-sphere smoothly
degenerates combining with the $\rho$ coordinate to form the origin
of $\mathbb{R}^{6}$. Third, we study the surface close to the conducting disks, i.e. $\eta\sim\eta_i$ and $0<\rho<R_i$ and show the 2-sphere combines with $\eta$ to smoothly describe the origin of $\mathbb{R}^{3}$. Finally, we study the region close to the rim of the disk, i.e. $\eta\sim\eta_i$ and $\rho\sim R_i$, and show its smoothness.

\paragraph{Regularity close to the linear charge distribution ($U(1)$ shrinking):} Consider the region $\rho\to 0$ close to the charge density \eqref{eq:glinear}. In this case we have
\begin{equation}
\dot{V} \to 2k\eta\,, \quad \quad
\dot{V}^{\prime} \to 2k\,, \quad \quad \ddot{V}\to 0\,,
\end{equation}
while $V^{\prime\prime}$ and $\dot{V}^{\prime}$ remain finite. The metric \eqref{eq:metric} is
\begin{align}
ds_{11}^{2}= & \left(-\frac{\dot{V}\Delta}{2V^{\prime\prime}}\right)^{\frac{1}{3}}\left[4\, d\Omega_{5}^{2}-\frac{2V^{\prime\prime}\dot{V}}{\Delta}\, d\Omega_{2}^{2}+ds_{4}^{2}\right] \\
ds_{4}^{2}= & -\frac{2V^{\prime\prime}}{\dot{V}}\left(d\rho^{2}+d\eta^{2}+\rho^{2}\,\frac{1}{4k^{2}}d\beta^{2}\right)-16\frac{k^{2}}{\Delta}\,\left(dt+v\right)^{2}\\
v= & \left(1-\frac{1}{2k}\right)\, d\beta\\
\Delta= & 2\dot{V}V^{\prime\prime}+\left(\dot{V}^{\prime}\right)^{2}\end{align}
which leads to a smooth combination\footnote{By definition, the periodicity of is $\beta\sim\beta +4k\pi$ for finite integer $k$. Thus, $\beta/(2k)$ has period $2\pi$ and consequently the $\rho\to 0$ limit describes the origin of $\mathbb{R}^{2}$ for any $k$.} of $\rho$ and $\beta$ to form $\mathbb{R}^{2}$. Of course, one needs to check that $V^{\prime\prime}<0$ and $\Delta>0$ for the above metric to be meaningful. 

\paragraph{Regularity at $\rho=0$, away from the linear charge distribution ($S^{5}$ shrinking):}

At $\eta<0$ and in the absence of sources, the potential $V$ admits
the series expansion
\begin{equation}
V= f_{0}\left(\eta\right)+f_{1}\left(\eta\right)\,\rho^{2}+f_{2}\left(\eta\right)\,\rho^{4} + \CO(\rho^6)
\label{eq:V_exp}
\end{equation}
The Laplace equation gives the relation
\begin{equation}
f_{0}^{\prime\prime}=-4f_{1}\,,
\label{eq:exp_relation}
\end{equation}
allowing us to write the scalar function $\Delta$ as
\begin{equation}
\Delta=4\rho^{4}\,\left[-2f_{2}f_{0}^{\prime\prime}+\left(f_{1}^{\prime}\right)^{2}\right]\equiv 4\rho^{4}\, g(\eta)\,.
\end{equation}
Hence the metric approaches
\begin{align}
ds_{11}^{2} & = g^{\frac{1}{3}}\,\left[4\rho^{2}\, d\Omega_{5}^{2}+4d\rho^{2}+\frac{\left(f_{0}^{\prime\prime}\right)^{2}}{g}\, d\Omega_{2}^{2}+4\left(d\eta^{2}+\frac{1}{\left(f_{0}^{\prime\prime}\right)^{2}}\, d\beta^{2}\right)\right. \nonumber \\
& \left. -\frac{\left(f_{0}^{\prime\prime}\right)^{2}}{g}\,\left(dt+\left(1-\frac{2f_{1}^{\prime}}{\left(f_{0}^{\prime\prime}\right)^{2}}\right)\, d\beta\right)^{2}\right]\,.
\end{align}
Notice for the latter to make sense one must check that 
\begin{equation}
\left(f_0^{\prime\prime\prime}\right)^2 > 2f_2\,f_0^{\prime\prime}\,\,(g>0) \quad \text{and} \quad f_1\neq 0\,.
\end{equation}
It is explicit that the 5-sphere shrinks to zero size smoothly by combining with $\rho$ to describe the origin of $\mathbb{R}^{6}$.

\paragraph{Regularity on the disk surfaces ($S^{2}$ shrinking):}

In this subsection we consider the area $\rho > 0$ close to one of the conducting
disks, i.e. $\eta\sim\eta_{i}$. Without loss
of generality we will take $\eta>\eta_{i}$. Since there is some induced
surface charge density on the disk, the potential $V$ admits the expansion
\begin{equation}
V=V_{0}+V_{1}\left(\rho\right)\,\left(\eta-\eta_{i}\right)+\frac{1}{6}V_{2}\left(\rho\right)\,\left(\eta-\eta_{i}\right)^{3}+\cdots\end{equation}
leading to the metric \eqref{eq:metric}
\begin{align}
ds_{11}^{2} & =\left(-\frac{\dot{V}_{1}^{3}}{2V_{2}}\right)^{\frac{1}{3}}\left[4d\Omega_{5}^{2}-2\frac{V_{2}}{\dot{V}_{1}}\left(\tilde{\eta}^{2}\, d\Omega_{2}^{2}+d\tilde{\eta}^{2}+d\rho^{2}+\frac{\rho^{2}}{\dot{V}_{1}^{2}}\, d\beta^{2}\right)-4\,\left(dt+v\right)^{2}\right]\,\\
v & =\left(1-\frac{1}{\dot{V}_{1}}\right)\, d\beta\,,
\end{align}
where $\tilde{\eta}=\eta-\eta_i$. The above shows how the 2-sphere combines with $\tilde{\eta}$ to smoothly form
the origin of $\mathbb{R}^{3}$.

\paragraph{Regularity close to the rim of a conducting disk:} If we examine the region very close to the rim of a disk of radius $R_i$ locate at $\eta_i$, we can forget about its curvature. The electrostatic problem is equivalent to that of a semi-infinite conducting plane with some boundary conditions. First, the induced charge distribution close to its edge vanishes. Second, there exists a discontinuity in the first derivative $\partial_{\eta}V$ at $\eta=\eta_{i}$ and $\rho<R_{i}$. Third, the electric field $\partial_{\rho}V$ vanishes on the semi-infinite plane. The electrostatic potential $V$ satisfying all these conditions is
\begin{align}\label{eq:V_close_to_rim}
\partial_{\eta}V&=\frac{q_{i}}{2}\,\frac{\eta-\eta_{i}}{\left|\eta-\eta_{i}\right|}\,\left(R_{i}-\rho+\sqrt{\left(\eta-\eta_{i}\right)^{2}+\left(\rho-R_{i} \right)^{2}} \right)^{1/2}+k_{i}+\cdots\\
\partial_{\rho}V&=-\frac{q_{i}}{2}\,\left(\rho-R_{i}+\sqrt{\left(\eta-\eta_{i} \right)^2+\left(\rho-R_{i}\right)^2}\right)^{1/2}+\cdots
\end{align}
for some constants $\{q_{i}\,,k_{i}\}$. Using these along with the change of coordinates
\begin{align}
\rho=R_{i}+\frac{1}{2}R_{i}\,\left(x^{2}-y^{2} \right)\nonumber \\
\eta=\eta_{i}+R_{i}\,xy,\quad x>0\,,
\end{align}
the metric \eqref{eq:metric} now reads
\begin{equation}
q_i^{-2/3}\,ds_{11}^{2}=2R_{i}\,d\Omega_{5}^{2}+R_{i}\left[dx^{2}+x^{2}\,d\Omega_{2}^{2}\right]+R_{i}\,dy^{2}+4q_i^{-2}\,d\beta^{2}
-2R_{i}\,\left(dt+d\beta\right)^{2}
\end{equation}
which is manifestly regular.

\subsection{Flux quantisation}

quantisation of fluxes requires the identification of smooth four- and seven-cycles in the geometries. In general,  we can think of an $n+2$ dimensional cycle as a disk fibration over an $n$ dimensional cycle. For example, consider the metric for an $n+2$ sphere
\begin{equation}
ds^{2}_{n+2}=\sin^{2}\theta\,d\phi^{2}+d\theta^{2}+\cos^{2}\theta\,d\Omega_{n}^{2},\quad 0<\theta<\pi/2
\end{equation}
$\theta=0$ describes the center of the fibered disk with an $S^{1}$ smoothly collapsing while at the rim of the disk $\theta=\pi/2$ it is an $n$ sphere that collapses smoothly.

Thus, the loci where lower dimensional cycles collapse smoothly in our geometries are natural candidates to construct our four- and seven-cycles. Our regularity analysis showed that 2-spheres smoothly collapse on the surface of the conducting disks, whereas 5-spheres do at the segments between the disks. Furthermore at the semi-infinite charge line distribution, an $S^1$ is smoothly collapsing too. These are the ingredients we will use to construct our smooth cycles.
\begin{figure}
\centering
\subfloat[A four-cycle]{\includegraphics[scale=0.5]{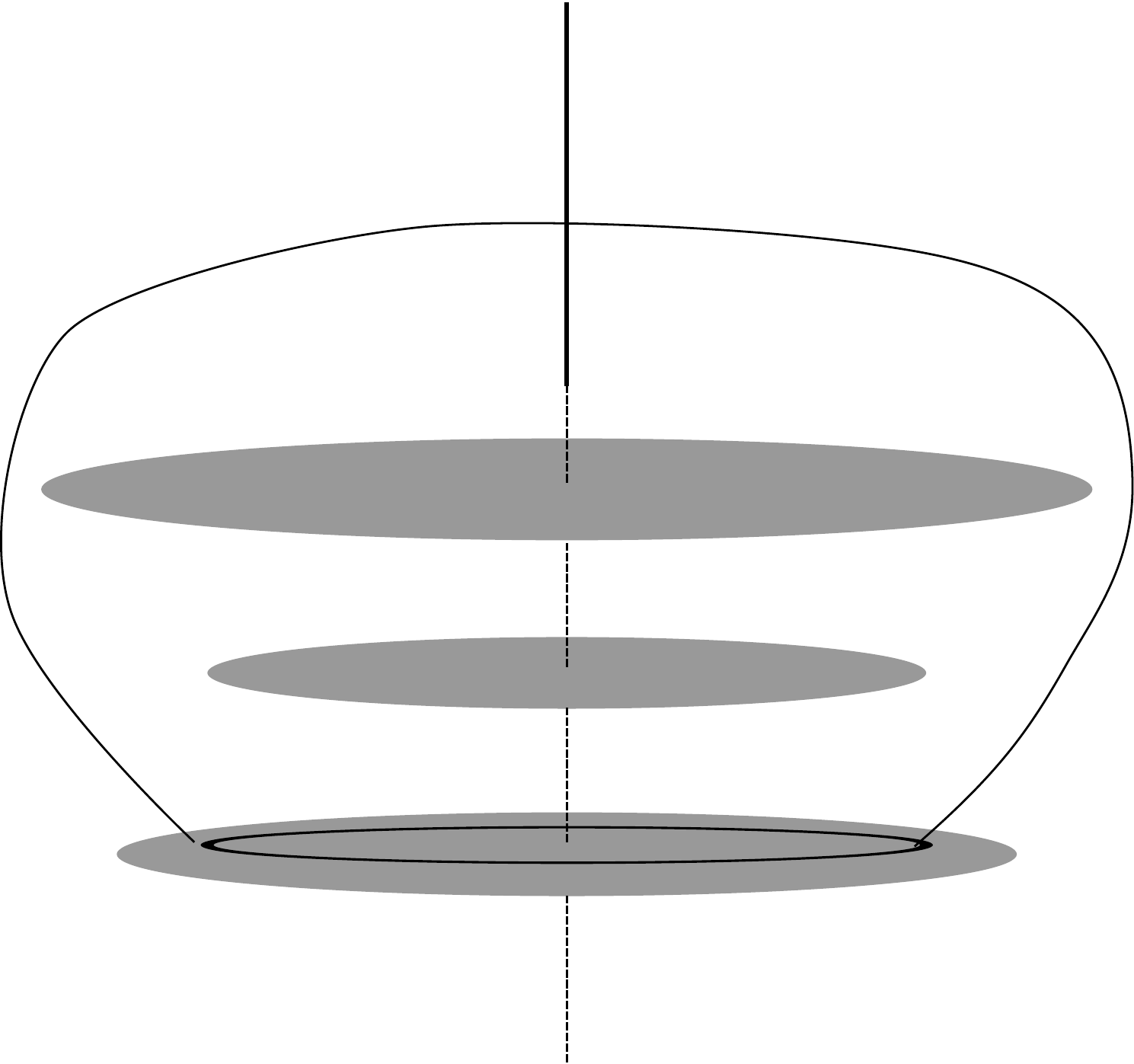}}
\qquad
\subfloat[A seven-cycle]{\includegraphics[scale=0.5]{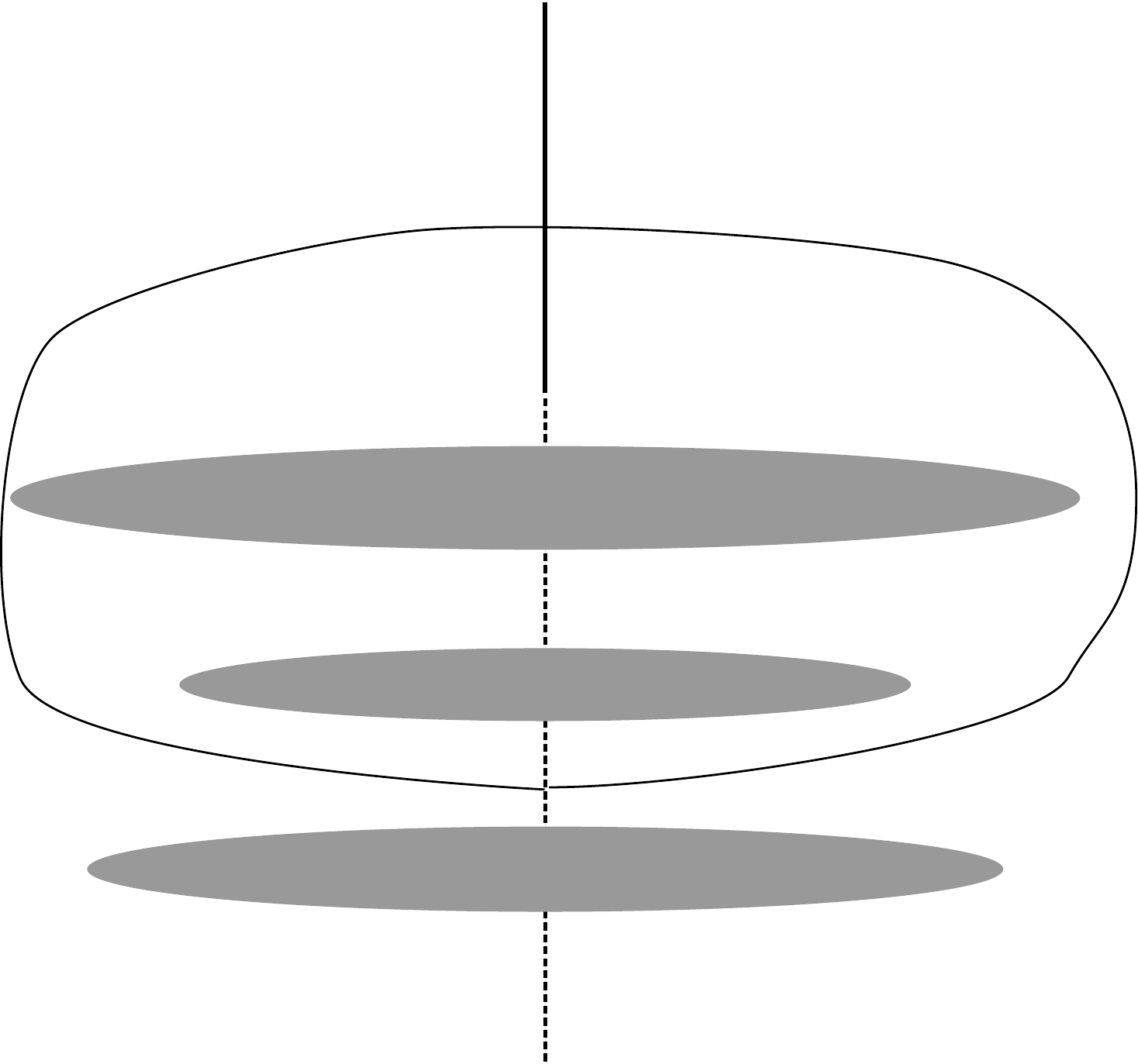}}
\caption{Surfaces describing 4-cycles and 7-cycles giving rise to non-trivial fluxes.}
\label{4-7cycles}
\end{figure}

More specifically, to construct a four cycle, we consider a two dimensional surface inside the three dimensional space spanned by $\rho$, $\eta$ and $\beta/2k$ as shown in figure \ref{4-7cycles}. The two dimensional surface we consider is such that it intersects with the linear charge distribution once where an $S^{1}$ collapses and it ends on an $S^{1}$ on one of our conducting disks where an $S^{2}$ collapses.

To construct a seven-cycle we consider a closed two dimensional surface which doesn't intersect with any conducting disk as shown in figure \ref{4-7cycles} but it surrounds a number of them. The surface, as shown,  intersects the $\rho=0$ line at one point on the line charge  distribution where again an $S^{1}$ smoothly shrinks and at a point on a line segment between two disks where an $S^{5}$ smoothly degenerates.

\begin{figure}
\centering
\includegraphics[scale=0.6]{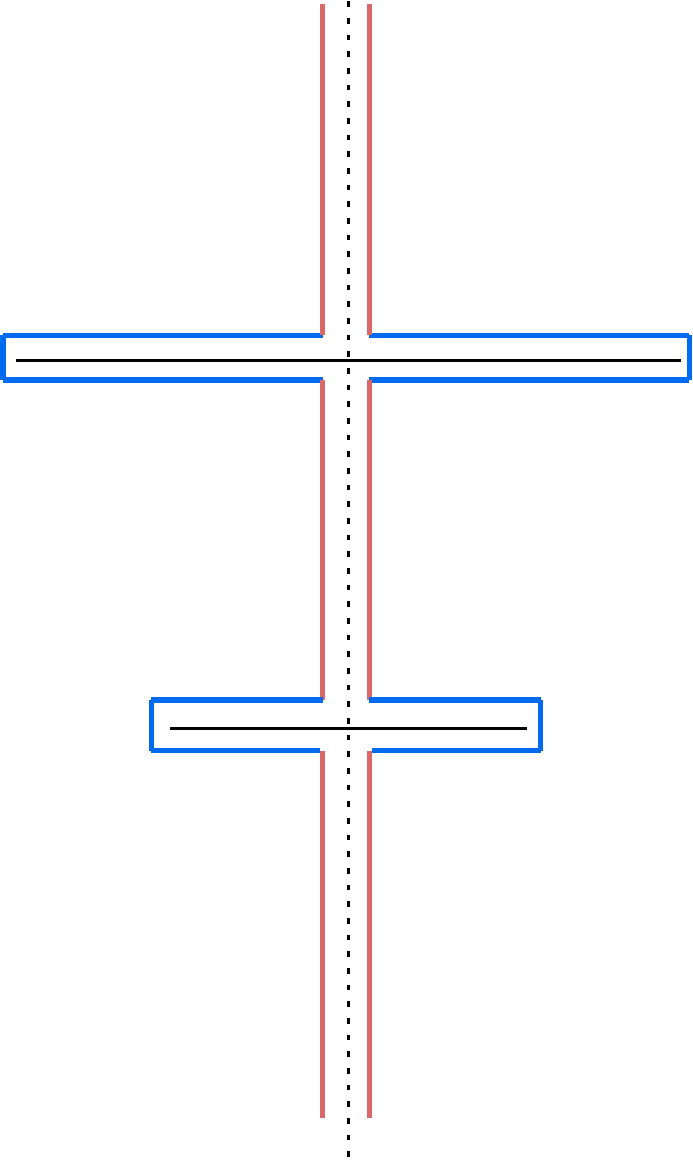}
\caption{Deformed surface used to construct four- and seven-cycles. The seven-form integral receives non-zero contribution only from the blue segments while the four-form integral receives non-zero contribution from the red line segments.}
\label{Cin}
\end{figure}

In order to quantize the fluxes of our solutions we need to integrate the four form over the four cycles and its hodge dual seven-form over the seven-cycles using the quantisation conditions
\begin{align}
N_{2} & =\frac{1}{16\pi G_{11}T_{M_{2}}}\int_{C_{7}}\ast_{11}G_{4}\label{N2_integral}\\
N_{5} & =\frac{1}{16\pi G_{11}T_{M_{5}}}\int_{C_{4}}G_{4}\label{N5_integral}
\end{align}
Since the integrated four and seven forms are regular, we can continuously deform the surfaces close to the $\rho=0$ axis and the disks without changing the value of the integral. Observe that  \eqref{N2_integral} only receives non-zero contribution from the surfaces of the disks while \eqref{N5_integral} only does from the $\rho=0$ line segments as shown in figure \ref{Cin}.

We conclude that each disk contributes a number of M2-branes
\begin{equation}
N_{2}^{i} =\frac{2\cdot2k}{\pi^{2}\ell_{p}^{6}}\int_{0}^{R_{i}}\rho^{2}\partial_{\rho}\left(\partial_{\eta}V_{\eta\rightarrow\eta_{i}^{+}}-\partial_{\eta}V_{\eta\rightarrow\eta_{i}^{-}}\right)d\rho=-\frac{2\cdot2k}{\pi^{3}\ell_{p}^{6}}Q_{i}
\end{equation}
while each segment between two consecutive disks contributes a number of M5-branes
\begin{equation}
N_{5}^{i} =\frac{2k}{\pi \ell_{p}^{3}}\left.\int_{d_{i+1}}^{d_{i}}\frac{\rho}{\partial_{\rho}V}\partial_{\eta}^{2}V\right|_{\rho=0}=\frac{4k}{\pi \ell_{p}^{3}}\left(d_{i}-d_{i-1}\right)\,.
\end{equation}
These arguments also apply for the vacuum configurations. In particular, notice the AdS${}_7\times$S$^{}_4$ does not carry M2-brane flux despite having a non-vanishing charge density because there is no smooth finite size 4-cycle available.

\subsection{Excited states: the conducting disk problem}

We want to construct more general solutions to \eqref{eq:Laplace} with suitable boundary conditions to be interpreted as excitations over AdS${}_4\times$S${}^7$ and AdS${}_7\times$S${}^4$. We first discuss the extension of the single disk configuration to an excited state. We then move to describe the mathematical problem determining a multiple disk configuration. 

\paragraph{Single conducting disk:} The single conducting disk is parameterised by its location $\eta_0$, its radius $R$ and the value of the potential on it $V_0$. AdS${}_4\times$S${}^7$ is a particular case where $\eta_0=0$, $V_0=\pi\,R/2$ and $R$ determines the radius of AdS${}_4$. From our cycle discussion, it is clear the reason why this single disk has no energy is because there is no 4-cycle giving rise to M5-brane flux. If we want to construct a configuration with non-vanishing energy all we have to do is to move down the location of the disk to an arbitrary $\eta_0 < 0$, keeping the background potential $V_b$. This is the configuration we study next.

The boundary problem is the same as for AdS${}_4\times$S${}^7$, since there is a single conducting disk, but the continuous and differentiable function appearing in Copson's solution now equals  $f(\rho)=V_{0}-V_{b}\left(\rho,\eta_{0}\right)$ where
\begin{equation}
  V_{b}\left(\rho,\,\eta_{0}\right)=\sqrt{\eta_{0}^{2}+\rho^{2}}+\eta_{0}\,\log\left[-\eta_{0}+\sqrt{\eta_{0}^{2}+\rho^{2}}\right]\,.
\end{equation}
Equation \eqref{eq:Copson_S} yields
\begin{equation}
  S_{\eta_{0}}\left(\rho\right)=\frac{V_{0}+k\eta_{0}}{2\pi}-\frac{k}{8\pi}\,\left[\pi\rho+2\rho\,\tan^{-1}\left(\frac{\eta_{0}}{2\rho}-\frac{\rho}{2\eta_{0}}\right)+2\eta_{0}\ln\left[4\left(\eta_{0}^{2}+\rho^{2}\right)\right]\right]\end{equation}
which reproduces  $S_{0}=\frac{V_{0}}{2\pi}-\frac{k}{4}\rho$ in the limit $\eta_{0}\rightarrow0^{-}$.

Using the above expression in \eqref{eq:Copson_sigma} we find an
integral over $\rho$ which cannot be easily evaluated but we will
write
\begin{equation}
\sigma_{\eta_{0}}=\int_{0}^{\eta_{0}}\partial_{\eta_{0}}\sigma_{\eta_{0}}+\sigma_{0}\end{equation}
yielding
\begin{align}
\sigma_{\eta_{0}}= & \frac{4}{\pi}\left(R\,\tan^{-1}\left(\frac{\eta_{0}}{R}\right)+\frac{\eta_{0}}{2}\ln\left[4\left(\eta_{0}^{2}+R^{2}\right)\right]-\frac{V_{0}}{k}-\eta_{0}+\frac{\pi}{2}R\right)\,\frac{1}{\sqrt{R^{2}-\rho^{2}}} \nonumber \\
 &-2\,\ln\left(\frac{R+\sqrt{R^{2}-\rho^{2}}}{\rho}\right)
 -\frac{4}{\pi}\,\int_{0}^{\eta_{0}}\frac{1}{\sqrt{\eta^{2}+\rho^{2}}}\tan^{-1}\left(\sqrt{\frac{R^{2}-\rho^{2}}{\eta^{2}+\rho^{2}}}\right)d\eta
 \end{align}
Requiring continuity in $\partial_\eta V$ at the rim of the disk $\rho=R$ fixes the disk potential $V_0$ to be
\begin{equation}
V_{0}=R\,\tan^{-1}\left(\frac{\eta_{0}}{R}\right)+\frac{\eta_{0}}{2}\ln\left[4\left(\eta_{0}^{2}+R^{2}\right)\right]-\eta_{0}+\frac{\pi}{2}R\,,
\end{equation}
since the last integral is continuous. Thus, after integration by parts, we can write the charge density as
\begin{align}
\sigma_{\eta_{0}}= & -2\,\log\left(R+\sqrt{R^{2}-\rho^{2}}\right)+2\log\rho\,\left(1-\frac{2}{\pi}\tan^{-1}\left(\frac{\sqrt{R^{2}-\rho^{2}}}{\rho}\right)\right)\nonumber \\
 & +\frac{4}{\pi}\log\left(-\eta_{0}+\sqrt{\eta_{0}^{2}+\rho^{2}}\right)\tan^{-1}\left(\sqrt{\frac{R^{2}-\rho^{2}}{\eta_{0}^{2}+\rho^{2}}}\right)\nonumber \\
 & +\frac{4\sqrt{R^{2}-\rho^{2}}}{\pi}\int_{0}^{\eta_{0}}\log\left(-\eta+\sqrt{\eta^{2}+\rho^{2}}\right)\frac{\eta}{\left(R^{2}+\eta^{2}\right)\sqrt{\left(\eta^{2}+\rho^{2}\right)}}\, d\eta\,.
 \end{align}

We can also evaluate the two non-trivial fluxes. The M5-brane flux equals
\begin{equation}
  N_5 = 2\sqrt{2N_2}\,\frac{-\eta_0}{L_4^3} \quad \Leftrightarrow \quad \frac{-2\eta_0}{L_4^3} = \frac{N_5}{\sqrt{2N_2}}\,.
\end{equation}
Thus, the location of the disk $\eta_0$ is fixed, in units of the AdS${}_4$ radius, in terms of the flux ratio $N_5/\sqrt{N_2}$. 
The total induced charge $Q(\eta_0,\,R)$ for any solution to the Copson electrostatic problem, using equations \eqref{eq:Copson_S} and \eqref{eq:Copson_sigma}, is given by
\begin{equation}
Q=2\pi\int_{0}^{R}\rho d\rho\,\sigma\left(\rho\right)=-16\pi\,\int_{0}^{R}S\left(\rho\right)d\rho=-8\,\int_{0}^{R}\frac{t}{\sqrt{R^{2}-t^{2}}}f\left(t\right)dt\,.
\end{equation}
For the current case, this equals
\begin{align}
Q= & 2\eta_{0}^{2}\pi-4\eta_{0}R-3\pi R^{2}+2R^{2}\tan^{-1}\left(\frac{\eta_{0}}{2R}-\frac{R}{2\eta_{0}}\right)\nonumber \\
 & +4\eta_{0}^{2}\tan^{-1}\left(\frac{\eta_{0}}{R}\right)+8\eta_{0}^{2}\tan^{-1}\left(\frac{R}{\eta_{0}}\right)-8R^{2}\tan^{-1}\left(\frac{\eta_{0}}{R}\right)\,.
 \end{align}
Since the solution is asymptotic to AdS${}_4\times$S${}^7$ with radius of curvature given by $L_4$, the total induced charge $Q$ must  satisfy
\begin{equation}
  -\frac{Q(\eta_0,\,R)}{L_4^6} = \frac{\pi}{2}\,,
\end{equation}
which determines the size of the disk $R$.

\paragraph{The multi-conducting disk problem:} The most general excited configuration in AdS${}_4\times$S${}^7$ will be characterized by a background potential $V_b$ associated with the semi-infinite line charge density $\lambda (\eta)$ and by a finite set of conducting disks located at $\eta=\eta_i$ with radia $R_i$. The background potential $V_b$ will induce some charge density $\sigma_i(\rho)$ on each of them, that will determine the disk potential $V_{d_i}$. The electrostatic potential solving our problem is
\begin{equation}
  V = V_b(\rho,\,\eta) + \sum_i V_{d_i}(\rho,\,\eta)\,,
\end{equation}
subject to the set of boundary conditions
\begin{equation}
  V_{0i} = V_b(\rho,\,\eta_i) + \sum_j V_{d_j} (\rho,\,\eta_i)\,.
 \label{eq:bproblem}
\end{equation}
There are as many boundary conditions as conducting disks. They provide a set of coupled integral equations for the charge densities $\sigma_i(\rho)$. We expect the constant values $V_{0i}$ will be fixed by requiring the continuity of the electric field $\partial_\eta V$ at the rim of each of the disks. The solution to such problem would provide the entire set of classical smooth configurations with AdS${}_4\times$S${}^7$ asymptotics.  

There is an extension of this problem for excited states over the AdS${}_7\times$S${}^4$ vacuum. In this case, there is also a background potential $V_b$ due to a semi-infinite line charge distribution and a collection of conducting disks. Using the method of images will take care of the boundary condition due to the infinite plane. Thus, the relevant electrostatic potential will be
\begin{equation}
  V =\begin{cases} V_b(\rho,\eta) - V_b(\rho,-\eta) + \sum_i \left(V_{d_i}(\rho,\,\eta)-V_{d_i}(\rho,\,-\eta)\right) + V_0, &\eta>0\\ V_{0},& \eta<0 \end{cases}\,.
\end{equation}
As before, there exist as many boundary conditions
\begin{equation}
  V_{0i} = V_b(\rho,\eta_i) - V_b(\rho,-\eta_i) + \sum_j \left(V_{d_j}(\rho,\,\eta_i)-V_{d_j}(\rho,\,-\eta_i)\right) + V_0\,,
\end{equation}
as finite conducting disks, giving rise to a coupled set of integral equations to determine the charge densities $\sigma_j$ on each disk.

\subsection{Mass for asymptotically $AdS_4\times S^7$ configurations \label{AdS4_mass}}

The general multi-ring configuration asymptotic to $AdS_4\times S^7$ is described by an electrostatic potential
\begin{align}
V= & V_b + \sum_i V_{d_i} = \sqrt{\eta^{2}+\rho^{2}}+\eta\,\log\left(-\eta+\sqrt{\eta^{2}+\rho^{2}}\right)
\nonumber \\
 & -\frac{1}{4\pi}\sum_{i}\int_{0}^{R_{i}}\rho^{\prime}d\rho^{\prime}\int_{0}^{2\pi}d\phi^{\prime}\,\frac{\sigma_{i}\left(\rho^{\prime}\right)}{\sqrt{\rho^{2}+\rho^{\prime2}-2\rho\rho^{\prime}\cos\phi^{\prime}+\left(\eta-\eta_{i}^{\prime}\right)^{2}}}\label{eq:full_potential}
 \end{align}
satisfying the boundary conditions \eqref{eq:bproblem}.

Even though we do not know the explicit mathematical solution to the boundary problem, one suspects the mass of the configuration should be determined by the first subdominant contribution in a multipole expansion of the full potential. Let us keep the lowest order terms in this expansion
\begin{equation}
V= \left[\sqrt{\eta^{2}+\rho^{2}}+\eta\,\log\left(-\eta+\sqrt{\eta^{2}+\rho^{2}}\right)\right]-\frac{F}{\left(\eta^{2}+\rho^{2}\right)^{\frac{1}{2}}}-P\,\frac{\eta}{\left(\eta^{2}+\rho^{2}\right)^{\frac{3}{2}}}+\cdots
\label{eq:ads4potexp}
\end{equation}
where we defined
\begin{align}
  F & =\frac{1}{4\pi}\sum_{i}Q_{i}= \frac{Q}{4\pi}\,, \label{eq:F_def}\\
  P & =\frac{1}{4\pi}\sum_{i}\eta_{i}Q_{i}=-\frac{1}{4\pi}\sum_{i}d_{i}Q_{i}=\frac{\pi^{3}l_{p}^{9}}{64}\sum_{i}\sum_{j=1}^{i}N_{2}^{j}N_{5}^{i}\,.
\label{eq:P_def}
\end{align}
$F$ is related to the total charge of the configuration and $P$ is the $\eta$ component of the dipole moment.

Instead of attempting a first principle calculation of the mass, we will match the asymptotic expansion of the general multi-ring configuration with the asymptotic expansion of the $AdS_4\times S^7$ half-BPS superstar. The latter has an eleven dimensional metric
\begin{align}
  ds_{11}^2 & = \Delta^{2/3}\left[-\frac{f_4}{H_4}\,dt^2 + \frac{dr^2}{f_4} + r^2\,d\Omega_2^2 + (2L_4)^2\,d\theta^2\right] \nonumber \\
  &+ \Delta^{-1/3}\,4L_4^2\left[\sin^2\theta\,d\Omega_5^2 + H_4\cos^2\theta\left(d\phi_1 + a_1\frac{dt}{2L_4}\right)^2\right]\,,
\label{eq:ads4superstar}
\end{align}
where $\Delta = 1+q_4\sin^2\theta/r$, $a_1=-q_4/(r+q_4)$, $H_4=1+q_4/r$ and $f_4=1+ r^2\,H_4/L_4^2$. Its asymptotic expansion at infinity equals
\begin{align}
ds_{11}^{2}= & ds_{2}^{2}+\left(1+\frac{2}{3}\frac{q_4}{r}\sin^{2}\vartheta\right)r^{2}d\Omega_{2}^{2}+4L_4^{2}\,\sin^{2}\vartheta\left(1-\frac{1}{3}\frac{q_4}{r}\sin^{2}\vartheta\right)d\Omega_{5}^{2} \nonumber \\
 & L_4^{2}\left(1+\frac{q_4}{r}\left(\frac{2}{3}\sin^{2}\vartheta-1\right)\right)\frac{dr^{2}}{r^{2}}+4L_4^{2}\,\left(1+\frac{q_4}{r}\frac{2}{3}\sin^{2}\vartheta\right)d\vartheta^{2}\,,
 \label{eq:bubble_expansion_2}
\end{align}
while its mass $M_4$ is given by \cite{Cai:2001jc}
\begin{equation}
  M_4 = \frac{q_4}{4G_4} = \frac{\left(2L_4\right)^{7}}{3\cdot2^{6}\cdot\pi^{3}\ell_{p}^{9}}\,q_4\,,
\label{eq:mass4}
\end{equation}
where we used $G_{4} =\frac{G_{11}}{V_{S^{7}}}=\frac{3}{\pi^{4}\left(2L_4\right)^{7}}G_{11}$ and
$G_{11} =16\pi^{7}\ell_{p}^{9}$. Since the mass is determined by the $AdS_4$ radius of curvature $L_4$ and the charge $q_4$, all we have to do is to match the asymptotics of the LLM metric with \eqref{eq:bubble_expansion_2}, and determine a dictionary between the multipole parameters $\{F,\,P\}$ and $\{L_4,\,q_4\}$.

Since the asymptotic LLM metric is $AdS_4\times S^7$, the natural coordinates to use in the asymptotic expansion of the full metric derived out of the potential expansion 
\eqref{eq:ads4potexp} are
\begin{equation}
\eta= -\frac{L_4^3}{2}\,R\cos 2\theta\,, \qquad \qquad
\rho= \frac{L_4^3}{2}\,\sqrt{1+R^2}\,\sin 2\theta\,, \quad
R\to \infty
\end{equation}
Comparing the lowest order contributions from both metrics determines $F$ as a function of $L_4$
\begin{equation}
  F= -\frac{L_4^{6}}{8}\,.
\label{eq:kF_constrain}
\end{equation}
Using this in the expansion gives
\begin{align}
ds_{11}^{2}= & \left(1-\frac{32P}{L_4^{9}R}\left(1+2\cos2\theta\right)\right)R^{2}d\Omega_{2}^{2}+4L_4^{2}\sin^{2}\theta\left(1+\frac{16P}{L_4^{9}R}\left(1+2\cos2\theta\right)\right)d\Omega_{5}^{2} \nonumber \\
 & L_4^{2}\left(1+\frac{16P}{L_4^{9}R}\left(1+2\cos2\theta\right)\right)\frac{dR^{2}}{R^{2}}+4L_4^{2}\left(1+\frac{16P}{L_4^{9}R}\left(1+2\cos2\theta\right)\right)d\theta^{2}\nonumber \\
&+\left(-4R^2+\frac{128 PR}{L_4^9}\left(1+2\cos2\theta \right) \right)\,\left(dt+
\frac{1}{2} d\beta \right)^2\\
&+\cos^{2}\theta\,\left(L_4^{2}+\frac{16\,P}{L_4^{7}R}\left(1+2\cos2\theta \right) \right)\,D\beta^2\nonumber \\
D\beta=&d\beta
-2\left(1-\frac{48P}{L_4^9 R} \right)\left(dt+\frac{1}{2} d\beta \right)\label{eq:4gauge_field}
\end{align}
To compare with \eqref{eq:bubble_expansion_2}, we need to consider both the further change of coordinates
\begin{equation}
R = \frac{r}{L_4} +48\frac{P}{L_4^{8}}\,\cos^{2}\vartheta\,, \quad \quad
\theta = \vartheta-12\frac{P}{L_4^{8}\,r}\,\sin2\vartheta\,.
\end{equation}
and the same identification of $\U(1)$'s as in \eqref{eq:time_scaling}, to read off the four dimensional gauge field $a_1$ correctly.
After these, we reproduce \eqref{eq:bubble_expansion_2} when the relation
\begin{equation}
 q_4=48\,\frac{P}{L_4^{8}}
\label{eq:P_asymptote}
\end{equation}
is satisfied. Thus, we conclude the mass of the multi-ring configuration asymptotic to AdS${}_4\times$S${}^7$ is given by 
\begin{equation}
  M_4 =\frac{1}{2L_4}\,\sum_{i}\sum_{j=1}^{i}N_{2}^{j}N_{5}^{i}\,.
\end{equation}
Even though we used a specific conformal frame to do our metric expansion matching, the result is conformally invariant since we proved all fluxes are invariant under the transformations \eqref{eq:conft}. This result reproduces the saturation of the BPS bound derived from superalgebra considerations 
\begin{equation}
  \Delta_4 \equiv M_4\,L_4 = \frac{1}{2}\,J\,,
\end{equation}
where $J$ stands for the $\U(1)$ R-charge. The factor of $1/2$ is consistent with the conformal dimension of an scalar field in three dimensions. This result is consistent with the microscopic considerations following the analysis of the moduli space of half-BPS configurations \cite{david,shahin-joan} in ABJM \cite{ABJM}. It is also consistent with the matrix model dual descriptions \cite{BMN} available in the limit $N\to\infty$, though our derivation is valid for finite $N$.

\subsection{Mass for asymptotically $AdS_{7}\times S^{4}$ configurations \label{AdS7_mass}}

We follow the same strategy here as in the previous section. First, we expand the total electrostatic potential
\begin{equation}
V\left(\rho,\eta\right)=
\begin{cases}
V_{b}\left(\rho,\eta\right)-V_{b}\left(\rho,-\eta\right)+V_{d}\left(\rho,\eta\right)-V_{d}\left(\rho,-\eta\right) + V_0\,, & \eta>0 \\
V_0\,, & \eta < 0 \end{cases}
\label{eq:full_potential_7}
\end{equation}
where the background potential $V_b(\eta,\rho)$ is given by \eqref{eq:ads7backpot} and
\begin{equation}
V_{d}\left(\eta,\rho\right) =-\frac{1}{4\pi}\sum_{i}\int_{0}^{R_{i}}\int_{0}^{2\pi}\rho^{\prime}d\rho^{\prime}d\phi^\prime\,\frac{\sigma_{i}\left(\rho^{\prime}\right)}{\left(\rho^{\prime2}+\rho^{2}-2\rho^{\prime}\rho\cos\phi^\prime+\left(\eta-\eta_{i}\right)^{2}\right)^{\frac{1}{2}}}\label{eq:disc_potential}
\end{equation}
corresponds to the sum of the potentials due to each disk located at $\eta=\eta_i > 0$, with surface charge density $\sigma_i$ solving the integral equation
\begin{equation}
V\left(\rho,\eta_{i}\right)=V_{i},\quad\rho<R_{i}
\end{equation}
and vanishing on the disk rim.

When performing the multipole expansion, there will be no point charge contribution due to the cancellation caused by the mirror charge. Thus the dominant contribution to the potential will be the dipole term
\begin{align}
V\left(\rho,\eta\right) & =V_{b}\left(\rho,\eta\right)-V_{b}\left(\rho,-\eta\right)-2P\,\frac{\eta}{\left(\eta^{2}+\rho^{2}\right)^{\frac{3}{2}}}+\cdots\\
P & =\frac{1}{4\pi}\sum_{i}\eta_{i}Q_{i}=-\frac{\pi^{3}l_{p}^{9}}{2^{4}}\sum_{i}\sum_{j=1}^{i}N_{2}^{j}N_{5}^{i}\label{eq:dipole_moment_7}\\
Q_{i} & =2\pi\int_{0}^{R_{i}}\rho d\rho\,\sigma_{i}
\end{align}
where we used the quantisation conditions \eqref{eq:M2_number} and
\eqref{eq:M5_number}. Notice the flip of sign in the value of such dipole. This is due to the location of the disks being at positive $\eta_i$.

Instead of attempting a first principle calculation of the mass, we will match the asymptotic expansion of the general multi-ring configuration with the asymptotic expansion of the $AdS_7\times S^4$ half-BPS superstar. The latter has an eleven dimensional metric
\begin{align}
  ds_{11}^2 &= \tilde{\Delta}^{1/3}\left[-\frac{f_{7}}{H_7}\,dt^2 +\frac{dr^2}{f_{7}} + r^2\,d\Omega_5^2 + \frac{L_7^2}{4}\,d\vartheta^2\right] \nonumber \\
  & + \frac{L_7^2}{4}\,\tilde{\Delta}^{-2/3}\left[\sin^2\vartheta\,d\Omega_2^2 + H_7\cos^2\vartheta\,\left(d\phi_1+a_1\frac{dt}{L_7}\right)^2\right]\,,
\label{eq:ads7superstar}  
\end{align}
with $\tilde{\Delta}=1+q_7\sin^2\vartheta/r^4$, $H_7=1+q_7/r^4$, $a_1=2H_7^{-1}$ and $f_7=1+r^2H_7/L_7^2$. Its asymptotic expansion at infinity equals
\begin{align}
ds_{11}^{2}= & ds_{2}^{2}+r^{2}\left(1+\frac{q_7}{3r^{4}}\,\sin^{2}\vartheta\right)\, d\Omega_{5}^{2}+\frac{L_7^{2}}{4}\left(1+\frac{q_7}{3r^{4}}\,\sin^{2}\vartheta\right)d\vartheta^{2}\nonumber \\
 & +\frac{L_7^{2}}{4}\sin^{2}\vartheta\left(1-\frac{2q_7}{3r^{4}}\sin^{2}\vartheta\right)\, d\Omega_{2}^{2}+\left(\frac{L_7^{2}}{r^{2}}-\frac{L_7^{4}}{r^{4}}+\frac{L_7^{6}}{r^{6}}+\frac{4q_7}{3r^{6}}\left(\sin^{2}\vartheta-3\right)\right)\, dR^{2}\,,
 \label{eq:bubble7_exp}
 \end{align}
 where $ds_2^2$ stands for the two dimensional metric in the $t-\phi$ plane. Its mass $M_7$ is given by \cite{Cai:2001jc}
\begin{equation}
M=\frac{\pi^{2}}{4G_{7}}\,q_7= \frac{L_7^{4}}{3\cdot 2^7\pi^{3}\ell_{p}^{9}}\,q_7
\label{eq:mass7}
\end{equation}
where we used $G_{7}=  \frac{G_{11}}{V_{S^{4}}}=\frac{6}{\pi^{2}L_7^{4}}G_{11}$ and
$G_{11}= 16\pi^{7}\ell_{p}^{9}$. Since the mass is determined by the $AdS_7$ radius of curvature $L_7$ and the charge $q_7$, all we have to do is to match the asymptotics of the LLM metric with \eqref{eq:bubble7_exp}, and determine a dictionary between the multipole parameters $P$ and $\{L_7,\,q_7\}$.

Since the asymptotic LLM metric is $AdS_7\times S^4$, the natural coordinates to use in the asymptotic expansion of the full metric are
\begin{equation}
\rho =\frac{L_7^{3}}{8}\,R\sqrt{1+ R^{2}}\,\cos\theta\,, \quad \quad
\eta =\frac{L_7^{3}}{8}\left(\frac{1}{2}+R^{2}\right)\sin\theta\,, \quad \quad R\to \infty
\end{equation}
Performing the expansion, the metric is 
\begin{align}
ds_{11}^{2}= & L_7^{2}\left[R^{2}\left(1+\frac{2^{10}\,P}{L_7^{9}R^{4}}\left(1+5\cos\left(2\theta\right)\right)\right)\, d\Omega_{5}^{2}+\frac{\sin^{2}\theta}{4}\left(1-\frac{2^{11}\,P}{L_7^{9}R^{4}}\left(1+5\cos\left(2\theta\right)\right)\right)\, d\Omega_{2}^{2}\right]\nonumber \\
 & +\frac{L_7^{2}}{4}\left(1-\frac{2^{11}\,P}{L_7^{9}R^{4}}\left(1+5\cos\left(2\theta\right)\right)\right)\left[d\theta^{2}+\frac{1}{R^{2}}\left(1-\frac{1}{R^{2}}+\frac{1}{R^{4}}\right)\, dR^{2}\right] \nonumber \\
&+\left(-L_7^{2}R^{2}-L_7^{2}-\frac{2^{10}\,P}{L_7^{7}R^{2}}\left(1+5\cos2\theta\right) \right)\,dt^{2}\\
&+\cos^{2}\theta\left(\frac{L_7^{2}}{4}-\frac{2^{11}\,P}{L_7^{7}R^{4}}\,\left(1+5\cos2\theta\right) \right)\,D\beta^{2}\nonumber \\
D\beta=&d\beta+2\left(1+\frac{3\cdot 2^{12}\,P}{L_7^{9}R^4} \right)dt\,. 
\label{eq:gauge_connection7}
\end{align}
To match the asymptotic metric \eqref{eq:bubble7_exp} we perform the further change of coordinates 
\begin{equation}
R\,L_7 = r+\frac{q_{7}}{4r^{3}}\cos^{2}\vartheta\, \quad \quad
\theta =\vartheta-\frac{q_{7}}{4r^{4}}\sin2\vartheta\,,
\end{equation}
and identify the $\U(1)$'s as in \eqref{eq:matchad7} but in appropriate units, i.e. $d\beta=d\phi-\frac{2}{L_{7}}d\tau, \,dt=\frac{d\tau}{L_{7}}$. This allows us to identify the relation
\begin{equation}\label{eq:asymptotic_identification_7}
q_{7}=-\frac{3\cdot 2^{12}P}{L_7^{5}}\,,
\end{equation}
from which we conclude the mass $M_7$ is 
\begin{equation}
  M_7 = \frac{2}{L_7}\sum_{i}\sum_{j=1}^{i}N_{2}^{j}N_{5}^{i}\,.
\label{eq:M5_mass}
\end{equation}
This reproduces the saturation of the BPS bound derived from superalgebra considerations 
\begin{equation}
  \Delta_7 \equiv M_7\,L_7 = 2J\,,
\end{equation}
where $J$ stands for the $\U(1)$ R-charge. The factor of 2 is consistent with the conformal dimension of an scalar field in six dimensions. Our derivation at finite $N$ is consistent with the matrix model dual descriptions \cite{BMN} available in the limit $N\to\infty$.

%%%%%%% end of regular gravity section %%%%%%%%%%%%%%%%%%%

\section{Engineering superstar-like singular configurations}

Regular configurations are characterised by a finite set of discrete conducting disks satisfying: first, the electrostatic potential is some fixed constant on each disk; second, the charge density vanishes on the rim of each disk, so that $\partial_\eta V$ is continuous there; and third, $\dot{V}$ vanishes on each disk. When the distance between conducting disks is sent to zero, one is dealing with a continuous distribution whose boundary shape is described by a curve $R=R(\eta)$ determining the radius of the given disk in the limit, i.e. $0\leq\rho\leq R(\eta)$, for the disk located at $\eta$. We can infer what the boundary conditions should be using electrostatics. $\dot{V}$ must vanish inside the conductor, whereas $\partial_\eta V$ must remain continuous across its boundary. Notice the charge distribution is no longer localised at discrete values of $\eta=\eta_i$, but becomes an $\eta$ varying function. Indeed, denoting such charge distribution by $d(\eta,\rho)$, using Laplace's equation and the conditions described above, we conclude $\partial_\rho d=0$ and consequently
\begin{equation}
  d(\eta) = \partial_\eta^2 V_d\,,
\end{equation}
where $V_d$ is the potential on the disk. 

The vanishing of $\dot{V}$ at the boundary of the continuous conductor is responsible for the appearance of a singularity. Indeed, the metric \eqref{eq:metric} will be singular because both the round 5-sphere and 2-sphere will shrink to zero size in a non-smooth way, the time coordinate will become null while the $\eta$, $\rho$ metric components will blow up. The connection between this kind of singularities and continuous distributions of conducting disks was already emphasised in \cite{mark-coarsegraining}. It is interesting that the emergence of this kind of singularity has its origin in losing the information about the precise distribution of discrete disks. This is reminiscent of the physics reported in \cite{library} and motivated most of the comments in \cite{mark-coarsegraining}.

In this section, we will first embed the half-BPS AdS${}_4\times$S${}^7$ and AdS${}_7\times$ S${}^4$ superstars \cite{sugraref} in the LLM ansatz. We will describe the electrostatic set-up responsible for these configurations and identify the shape of the boundary of the continuous disk distribution responsible for their singularities. We will check our interpretations by computing the charges of both configurations using our derived charge densities $d(\eta)$. Having understood these particular cases, we will describe the boundary problem describing the most general singular configuration of this kind in terms of an arbitrary boundary curve $R(\eta)$.

\subsection{AdS${}_4\times$S${}^7$ superstar\label{sec:AdS4_superstar}}

The asymptotically $AdS_4\times S^7$ half-BPS superstar is an eleven dimensional supergravity configuration with metric given in \eqref{eq:ads4superstar}. The solution depends on two parameters : $L_4$ the radius of $AdS_4$ and $q_4$ which determines its mass and R-charge through \eqref{eq:mass4}.  Its microscopic interpretation is in terms of a distribution of giant gravitons over the 7-sphere \cite{Myers:2001aq}. The latter is consistent with the flux quantisation condition
\begin{equation}
  \frac{q_4}{L_{4}}=\frac{N_5}{\sqrt{2N_2}}\,,
\label{eq:ads4myers}
\end{equation}
providing the number of M5-branes supporting this configuration.

\paragraph{Matching LLM:} First, we embed \eqref{eq:ads4superstar} in
the class \eqref{eq:metric}. Focusing on the
subspace spanned by the transverse 5-sphere, the 2-sphere in $AdS_4$ and the $r-\vartheta$ plane, we can rewrite \eqref{eq:ads4superstar} as
\begin{equation}
ds_{11}^{2} =ds_{2}^{2}+\frac{\left(2L_4\right)^{2}\sin^{2}\vartheta}{4\Delta^{\frac{1}{3}}}\left[\frac{4\Delta r^{2}}{(2L_4)^{2}\sin^{2}\vartheta}\, d\Omega_{2}^{2}+4\, d\Omega_{5}^{2}+\frac{4\Delta}{\sin^{2}\vartheta}\,\left(\frac{dr^{2}}{\left(2L_4\right)^{2}f_4}+d\vartheta^{2}\right)\right]\,,
\label{eq:super_exp}
\end{equation}
where $ds_2^2$ corresponds to the 2-dimensional metric in the $t-\phi_1$ plane. To match both metrics, we consider the coordinate transformation
\begin{equation}
\rho= \frac{1}{2}\, w\left(r\right)\sin\left(2\vartheta\right)\,, \quad\quad \eta= -\frac{1}{2}\, h\left(r\right)\cos\left(2\vartheta\right)+s
\label{eq:trans4}
\end{equation}
for some constant $s$. Requiring the cross-term metric component $dr\,d\vartheta$ to vanish fixes
\begin{equation}
w(r)=\sqrt{h(r)^{2}+c}
\end{equation}
where $c$ is a constant of integration. On the other hand, the $\rho-\eta$ conformal metric in \eqref{eq:metric} satisfies
\begin{equation}
d\rho^{2}+d\eta^{2}=\left(h^{2}+c\cos^{2}\left(2\vartheta\right)\right)\left(d\vartheta^{2}+\frac{1}{4}\frac{h^{\prime^{2}}}{h^{2}+c}\: dr^{2}\right)\,.
\end{equation}
Since the ratio of the two metric components equals $(2L_4)^{2}\,f_4$, we derive 
\begin{equation}
h(r)=L_4^2\,\left(r+\frac{q_4}{2}\right)\,, \quad \quad
w(r)=L_4^2\,\left(r^{2}+rq_4+L_4^2\right)^{\frac{1}{2}}\,.
\end{equation}
Thus, the coordinate transformation \eqref{eq:trans4} is
\begin{align}
\rho= & \frac{L_4^{2}}{2}\,\left(r^{2}+rq_4+L_4^{2}\right)^{\frac{1}{2}}\sin\left(2\vartheta\right)\,,\\
\eta= & -\frac{L_4^{2}}{2}\,\left(r+\frac{q_4}{2}\right)\cos\left(2\vartheta\right)+s\,.
\end{align}
On the other hand, the 5-sphere and 2-sphere warp factors determine the
equation for the potential $V$
\begin{equation}
\dot{V}=\rho\partial_{\rho}V=L_4^{2}\, r\,\sin^{2}\vartheta\,.
\label{eq:y_super}
\end{equation}
The way to fix the constant $s$ is to use the asymptotics of the metric. The latter is AdS${}_4\times$S${}^7$ and we know it supports a semi-infinite line charge distribution $\lambda(\eta)=2\eta$ at $\rho=0$ for $\eta>0$, but vanishes for $\eta<0$. 
This location requires $\vartheta=\pi/2$. To make sure the line distribution ends at $\eta=0$, $s$ must equal $s=-q_4L_4^2/4$. Thus, the final coordinate transformation is
\begin{align}
\rho= & \frac{L_4^{2}}{2}\,\left(r^{2}+rq_4+L_4^{2}\right)^{\frac{1}{2}}\sin\left(2\vartheta\right)\,,\\
\eta= & -\frac{L_4^{2}}{2}\,\left(r+\frac{q_4}{2}\right)\cos\left(2\vartheta\right)-q_4\,\frac{L_4^2}{4}\,.
\end{align}

To interpret the conducting disk distribution supporting this configuration, we will examine the locations where $\dot{V}$ vanishes. Since $\vartheta\in [0,\,\frac{\pi}{2}]$, this occurs either for $\vartheta=0$ or $r=0$.
The $\vartheta=0$ region corresponds to $\rho=0$ and $\eta<-\frac{q_4\,L_4^{2}}{2}$. There is no charge density in this region because the potential is continuous. The $r=0$ region describes the ellipsis
\begin{equation}
\frac{\rho^{2}}{\left(2L_4\right)^{2}}+\frac{\left(\eta+\frac{q_4L_4^{2}}{4}\right)^{2}}{q_4^{2}}=\left(\frac{L_4^{2}}{4}\right)^{2}\,.
\label{eq:M2_ellipsis}
\end{equation}
This can be interpreted as a continuous distribution of conducting disks at locations $\eta$ in the interval  $\eta\in \left[-\frac{q_4L_4^{2}}{2},\ 0\right]$ with radii $R(\eta)$ determined by
\begin{equation}
\frac{R(\eta)^{2}}{\left(2L_4\right)^{2}}+\frac{\left(\eta+\frac{q_4L_4^{2}}{4}\right)^{2}}{q_4^{2}}=\left(\frac{L_4^{2}}{4}\right)^{2}\,.
\end{equation}
We will test this interpretation below, when we compute the charge density for this continuous distribution and the charges carried by the configuration.

To finish the matching, we must determine $\partial_{\eta}V$. This can be achieved by determining $x\left(r,\vartheta\right)$. Consider the $x-y$ plane. The $y$ coordinate is already determined by \eqref{eq:y_super} since $y=\dot{V}$. Consider the ansatz
\begin{equation}
x=L_4^{2}\,\mu(r)\,\cos\vartheta\,.
\label{eq:x_super}
\end{equation}
Plugging in the metric \eqref{eq:Metric_ansatz} and demanding no
cross-term between $r$ and $\vartheta$ we obtain 
\begin{equation}\label{D_var_k}
e^{D}=\sin^{2}\vartheta\,\frac{2r}{\mu\mu^{\prime}}
\end{equation}
leading to
\begin{equation}
dy^{2}+e^{D}dx^{2}=4L^{4}\,\left(\frac{1}{4}\sin^{2}\left(2\vartheta\right)+r\frac{\mu}{2\mu^{\prime}}\sin^{4}\vartheta\right)\left(d\vartheta^{2}+\frac{1}{2r}\frac{\mu^{\prime}}{\mu}dr^{2}\right)
\end{equation}
and after comparing with \eqref{eq:super_exp} we have the equation
\begin{equation}
\left(\log\mu\right)^{\prime}=\frac{1}{2}\frac{r}{r^{2}+rq_4+L_4^{2}}
\label{eq:mu(r)}
\end{equation}
determines $\mu$. We see that we will have an integration constant
which represents the symmetry of \eqref{eq:Metric_ansatz} scaling
the $x$, $\beta$ and at the same time shifting $D$. On the other
hand the meaning of the constant $k$ is demonstrated in the the coordinates
transformation of $x$ and $y$, it represents the general conformal
symmetry of the metric \eqref{eq:Metric_ansatz} of the $x_{i}$ plane. 

\paragraph{Charge density:} the previous discussion determines the size of the disks in terms of their locations, but not their charges. Instead of solving the boundary value problem, we notice that the potential inside the ellipsis $V_{in}$ only depends on $\eta$. This is because its interior contains a distribution of conducting disks requiring
\begin{equation}
\partial_{\rho}V_{in}=0
\end{equation}
on the surface of each disk. Furthermore, the electric field
must be continuous across the surface of the ellipsoid. Thus,
\begin{equation}
\partial_{\eta}V_{in}\left(\eta\right)=\left.\partial_{\eta}V\left(\rho,\eta\right)\right|_{\rho=\rho\left(\eta\right)}
\end{equation}
where $\rho\left(\eta\right)$ satisfies \eqref{eq:M2_ellipsis}. Using  \eqref{eq:trans3} and reminding that the ellipsis is located at $r=0$ in the $r-\vartheta$ plane,
we have
\begin{align}
\partial_{\eta}V_{in} & =\log x\left(0,\vartheta\right)=\frac{1}{2}\,\log \cos^{2}\vartheta+\log\mu\left(0\right)\\
 & =\frac{1}{2}\,\log\left(-\eta\right)-\frac{1}{2}\,\log\left(\frac{L_4^{2}q_4}{2}\right)+\log\mu\left(0\right)
 \end{align}
where $\eta\in \left(-L_4^{2}q_4,0\right)$. We now use the Laplace
equation to calculate the three dimensional charge density $d(\eta)$ inside
the ellipsis
\begin{equation}
d(\eta)= \partial_{\eta}^{2}V_{in}+\frac{1}{\rho}\partial_{\rho}\left(\rho\partial_{\rho}V_{in}\right)
= \partial_{\eta}^{2}V_{in}=\frac{1}{2\eta}\label{eq:charge_density}
\end{equation}
The total charge density $q(\eta)$ carried by the
disk located at $\eta$ is
\begin{equation}
q\left(\eta\right)= 2\pi\,\int_{0}^{R\left(\eta\right)}d\left(\eta\right)\rho\: d\rho
= \pi d\left(\eta\right)R^{2}\left(\eta\right)
= -\frac{\pi \left(2L_4\right)^{2}}{2q_4^{2}}\left(L_4^{2}q_4+\eta\right)
\end{equation}

The total charge density $q(\eta)$ allows us to compute the total charge $Q$ and total dipole moment $P$ carried by the configuration. When integrated over $\eta$
\begin{equation}
Q=\int_{-\frac{L_4^{2}q_4}{2k}}^{0}q\left(\eta\right)d\eta=-\frac{\pi}{2} L_4^{6}\,,
\end{equation}
it reproduces the total number of M2-branes using the identity \eqref{eq:kF_constrain}
\begin{equation}
  \frac{Q}{4\pi} = - \frac{L_4^6}{8}\,,
\end{equation}
derived from our asymptotic analysis. When calculating the total dipole moment, using the continuous version of  \eqref{eq:P_def}
\begin{equation}
P=\frac{1}{4\pi}\int_{-\frac{L_4^{2}q_4}{2k}}^{0}\eta q\left(\eta\right)\, d\eta=\frac{L_4^{8}q_4}{12}\,,
\end{equation}
we again reproduce the asymptotic match \eqref{eq:P_asymptote}.

The total number of M5-brane is obtained from the quantisation condition \eqref{eq:M5_number}. This only depends on the range of the $\eta$ coordinate where there exists a non-trivial conducting disk distribution. Thus,
\begin{equation}
N_{5}=\frac{4}{\pi \ell_{p}^{3}}\,\sum_{i}\left(\eta_{i+1}-\eta_{i}\right)=\frac{2L_4^{3}}{\pi \ell_{p}^{3}}\,\frac{q_4}{L_4}\,.
\end{equation}
This is again consistent with the microscopic relation \eqref{eq:ads4myers}.

\subsection{AdS${}_7\times$S${}^4$ superstar}

The asymptotically AdS${}_4\times$S${}^7$ half-BPS superstar is an eleven dimensional supergravity configuration with metric given in \eqref{eq:ads7superstar}. The solution depends on two parameters : $L_7$ the radius of AdS${}_7$ and $q_7$ which determines its mass and R-charge through \eqref{eq:mass7}.  Its microscopic interpretation is in terms of a distribution of giant gravitons over the 4-sphere \cite{Myers:2001aq}. The latter is consistent with the flux quantisation condition
\begin{equation}
  \frac{q_7}{L_{7}^4} = \frac{N_2}{N_5^2}\,.
\label{eq:ads7myers}
\end{equation}
where $N_2$ stands for the number of giant gravitons (M2-branes).

\paragraph{Matching LLM:}  To embed \eqref{eq:ads7superstar} into \eqref{eq:metric}, we consider
the coordinate transformation
\begin{equation}
\eta= h\left(r\right)\,\sin\vartheta\,, \quad \quad
\rho= w\left(r\right)\,\cos\vartheta\,.
\end{equation}
Demanding no cross-term in $dr\, d\vartheta$, we obtain
\begin{equation}
w=\sqrt{h^{2}+c}\,,
\end{equation}
for some integration constant $c$. We now follow a similar procedure to the one described in the previous subsection. We consider the $\rho-\eta$ plane
\begin{equation}
d\rho^{2}+d\eta^{2}=\left(h^{2}\,\cos^{2}\vartheta+\left(h^{2}+c\right)\,\sin^{2}\vartheta\right)\left[d\vartheta^{2}+\frac{h^{\prime2}}{h^{2}+c}\, dr^{2}\right]\,,
\end{equation}
compare with the superstar metric \eqref{eq:ads7superstar}, we obtain the transformation
\begin{align}
\eta= & \frac{L_{7}}{8}\left(\frac{L_{7}^{2}}{2}+r^{2} \right)\,\sin\vartheta\,, \label{eq:coord_trans_1}\\
\rho= & \frac{L_7}{8}\left(r^{4}+r^{2}L_{7}^{2}+q_{7}\right)^{\frac{1}{2}}\,\cos\vartheta\,.
\label{eq:coord_trans_2}
\end{align}
Comparing the 5-sphere and 2-sphere sizes also fixes
\begin{equation}
\dot{V}=\frac{L_7}{2^3}r^{2}\sin\vartheta\,.
\label{eq:dot_V_7}
\end{equation}

To identify the conducting disk distribution supporting this configuration, we examine the locations where $\dot{V}$ vanishes. Since $\vartheta\in [0,\,\frac{\pi}{2}]$, this occurs either for $\vartheta=0$ or $r=0$. The $\vartheta=0$ region corresponds to
$\eta=0$ and $\rho>\frac{L_7\,q_7^{\frac{1}{2}}}{8}$. There is no charge distribution since the potential is continuous here. The $r=0$ region describes a hemiellipsoid
\begin{equation}
\frac{\eta^{2}}{\left(\frac{L_7^{3}}{16}\right)^{2}}+\frac{\rho^{2}}{\left(\frac{L_7q_7^{\frac{1}{2}}}{8}\right)^{2}}=1\,,\quad\eta>0
\end{equation}
while on the semi-infinite line $\rho=0$ $\eta>\frac{L_7^{3}}{16}$
we have the line charge distribution
\begin{equation}
\lambda=\eta-\frac{L_7^{3}}{16}
\end{equation}
exactly as in the case of the vacuum.

In total it looks like we densely placed an infinite sequence of circular
conducting disks centered at $\rho=0$ and at $0<\eta_{i}<\frac{L^{3}_{7}}{16}$
with radii $R_{i}$ satisfying
\begin{equation}
\frac{\eta_{i}^{2}}{\left(\frac{L^{3}_{7}}{16}\right)^{2}}+\frac{R_{i}^{2}}{\left(\frac{L_{7}q_{7}^{\frac{1}{2}}}{8}\right)^{2}}=1,\quad\eta>0
\end{equation}

Using a similar procedure to the one in the previous subsection we can identify the transformation to the LLM coordinates
\begin{equation}
x=\frac{L_{7}}{8}\mu\left(r\right)\,\cos\vartheta
\end{equation}
for some function $\mu\left(r\right)$. For the current case, we find we should identify
\begin{equation}
e^{D}=\frac{2r^{3}}{\mu\mu^{\prime}}
\end{equation}
with the function $\mu$ satisfying the differential equation
\begin{equation}
\frac{2}{r}\frac{\mu^{\prime}}{\mu}=\frac{4r^{2}}{L_{7}^{2}r^{2}+r^{4}+q_{7}}
\end{equation}

For similar reasons to the previous subsection the above identification fixes the electric field inside the hemiellipsoid to be
\begin{equation}
\partial_{\eta}V_{in}=\log x\left(0,\vartheta\right)=\log\left(\frac{L_{7}\mu\left(0\right)}{8} \right)+\frac{1}{2}\log\left(1-\frac{16^{2}}{L_{7}^{6}}\,\eta^{2} \right)
\end{equation}
from where we can compute the charge density
\begin{equation}
d(\eta)= \partial_{\eta}^{2}V_{in}+\frac{1}{\rho}\partial_{\rho}\left(\rho\partial_{\rho}V_{in}\right)
= \partial_{\eta}^{2}V_{in}=-\frac{16^{2}}{L_{7}^{6}}\,\frac{\eta}{1-\frac{16^{2}}{L_{7}^{6}}\,\eta^{2}} \label{eq:charge_density_7}
\end{equation}
Finally, we compute
\begin{align}
P=\frac{1}{4\pi}\,\int_{0}^{\frac{L^{3}_{7}}{16}}\pi R^{2}\left(\eta\right)d\left(\eta\right)\,d\eta=-\frac{L_{7}^{5}}{3\cdot 2^{12}}\,q_{7}\,,
\end{align}
which matches our asymptotic analysis \eqref{eq:asymptotic_identification_7}.

\subsection{General superstar-like singular configurations}

Given a continuous distribution of conducting disks determined by the value of the potential at the disks $V_{d}(\eta)$ and its boundary shape $R(\eta)$, it is natural to ask what singular configuration it corresponds to. Using the Laplace equation fixes the conductor charge density 
\begin{equation}
d(\eta)=\partial_{\eta}^{2}V_{d}(\eta)\,.
\end{equation}
Continuity of the electric field at the boundary of the distribution requires
\begin{equation}
\left.n^{m}\partial_{m}V(R(\eta),\,\eta)\right|_{boundary}-\left.n^{m}\partial_{m}V_{d}(\eta)\right|_{boundary}=0\,,
\label{eq:boundary_AdS4}
\end{equation}
where $n^m$ stands for a vector normal to such boundary. In general, this is a non-trivial integral equation relating the shape of the conductor $R(\eta)$ to its charge density $d(\eta)$.  

This boundary condition  determines the potential uniquely. To see this, define
\begin{equation}
  \Phi=V(\rho,\,\eta) -V_{d}(\eta)
\end{equation}
inside the boundary $(R(\eta),\eta)$, where $V(\rho,\,\eta)$ stands for the total potential.
Because of the Laplace equation and \eqref{eq:boundary_AdS4}, this function satisfies
\begin{equation}
\Box\Phi=0\,, \quad \quad \left.n^{m}\partial_{m}\Phi\right|_{boundary}=0\,.
\end{equation}
The solution to this boundary problem is unique
\begin{equation}
\Phi=0 \quad \Rightarrow \quad V=V_{d}\,,
\end{equation}
Thus, the potential inside the conductor equals the potential of the disk distribution, as it should. 
This establishes the existence of a precise correspondence between the space of singular configurations of the type we have discussed in this section and the shape of the continuous conductor boundary $R(\eta)$.

On the other hand, we know the potential outside of the conductor. If the solution is asymptotically AdS${}_4\times$S${}^7$, then it is given by
\begin{equation*}
V\left(\rho,\eta\right)= V_{b}\left(\rho,\eta\right)-\frac{1}{4\pi}\,\int_{\eta_{1}}^{\eta_{2}}d\eta^{\prime}\int_{0}^{R\left(\eta^{\prime}\right)}d\rho^{\prime}\rho^{\prime}\int_{0}^{2\pi}d\phi\,\frac{d\left(\eta^{\prime}\right)}{\sqrt{\rho^{2}+\rho^{\prime2}-2\rho\rho^{\prime}\cos\phi+\left(\eta-\eta^{\prime}\right)^{2}}}\,,
 \label{eq:Total_potential_AdS4}
\end{equation*}
where $0>\eta_2>\eta>\eta_1$ and $V_b$ stands for \eqref{eq:bpotential4}. If it is asymptotic to AdS${}_7\times$S${}^4$, then
\begin{align*}
V\left(\rho,\eta\right)= & V_{b}\left(\rho,\eta\right)-V_{b}\left(\rho,-\eta\right)+V_{dp}\left(\rho,\eta\right)-V_{dp}\left(\rho,-\eta\right)\label{eq:Total_potential_AdS7}\\
V_{dp}\left(\rho,\eta\right)= & -\frac{1}{4\pi}\,\int_{\eta_{1}}^{\eta_{2}}d\eta^{\prime}\int_{0}^{R\left(\eta^{\prime}\right)}d\rho^{\prime}\rho^{\prime}\int_{0}^{2\pi}d\phi\,\frac{\partial_{\eta}^{2}V_{d}\left(\eta^{\prime}\right)}{\sqrt{\rho^{2}+\rho^{\prime2}-2\rho\rho^{\prime}\cos\phi+\left(\eta-\eta^{\prime}\right)^{2}}}
\end{align*}
where $L_4^3/16>\eta_2>\eta>\eta_1>0$ and $V_b$ stands for \eqref{eq:ads7backpot}.

%%%%%%%%%%%%%  ACKNOWLEDGEMENTS  %%%%%%%%%%%%%%%%%%%%%

\section*{Acknowledgements}
We would like to thank Nadav Drukker, Iosif Bena and Nick Warner for discussions.
AD would like to thank the University of Edinburgh for hospitality during the intermediate stages of this project. The work of AD is supported by an EPSRC Postdoctoral Fellowship.
The work of JS was partially supported by the Engineering and
Physical Sciences Research Council [grant number EP/G007985/1].

%%%%%%%%%%%%% BIBLIOGRAPHY %%%%%%%%%%%%%%%%%%%%%%%%%%

\providecommand{\href}[2]{#2}\begingroup\raggedright

\endgroup

\end{document}